\RequirePackage{ifpdf}
\documentclass[12pt]{article}
\usepackage{epsfig,graphicx,bm,amsmath,amsfonts}
\usepackage{epstopdf}
\usepackage{amssymb,xcolor,float}
\usepackage{amsthm}
\usepackage{hyperref}
\usepackage{cite}
\input epsf.sty
\usepackage{cite}
\usepackage{comment}
\usepackage{epstopdf}
\usepackage{float}
\usepackage{caption}
\usepackage{subcaption}
\usepackage{enumitem}
\usepackage{bbm}
\usepackage{bm}
\usepackage{longtable}
\usepackage{parskip}
\usepackage{enumitem}
\usepackage{graphicx,epsfig}
\usepackage{ae,aecompl}
\usepackage{dcolumn}
\usepackage{latexsym}
\usepackage{psfrag}
\usepackage{ytableau}
\usepackage{tabularx}
\usepackage{float}
\usepackage{wrapfig}

\usepackage{xcolor}
\usepackage[english]{babel}
\usepackage[T1]{fontenc}
\usepackage[utf8]{inputenc}
\usepackage{authblk}
\usepackage{mathtools}
\usepackage{slashed}
\usepackage{mattens,amssymb}
\usepackage{amsmath,amssymb}
\usepackage{amsfonts}
\usepackage{graphicx,color}
\usepackage{cite}
\usepackage{hyperref}
\hypersetup{
	colorlinks=true,
	linkcolor=blue,
	filecolor=magenta,      
	urlcolor=cyan,
}
\usepackage[capitalize]{cleveref}
\numberwithin{equation}{section} 

\definecolor{refcol}{rgb}{0.9,0.1,0.1}
\hypersetup{colorlinks=true,linkcolor=blue,citecolor=refcol,urlcolor=cyan,linktocpage}

\newcommand{\Tr}{\text{Tr}}

\newcommand{\ben}{\begin{eqnarray}\displaystyle}
\newcommand{\een}{\end{eqnarray}}

\newcommand{\be}{\begin{equation}}
\newcommand{\ee}{\end{equation}}

\newcommand{\bc}{\begin{center}}
\newcommand{\ec}{\end{center}}

\newcommand{\eesp}{\end{split}}
\newcommand{\bsp}{\begin{split}}

\newcommand{\Rmnum}[1]{\expandafter\@slowromancap\romannumeral #1@}

\renewcommand{\l}{\lambda}	

\newcommand{\cB}{\mathcal{B}}
\newcommand{\cC}{\mathcal{C}}
\newcommand{\cD}{\mathcal{D}}

\newcommand{\cF}{\mathcal{F}}
\newcommand{\cG}{\mathcal{G}}
\newcommand{\cH}{\mathcal{H}}

\newcommand{\cK}{\mathcal{K}}
\newcommand{\cL}{\mathcal{L}}
\newcommand{\cM}{\mathcal{M}}
\newcommand{\cN}{\mathcal{N}}

\newcommand{\cR}{\mathcal{R}}
\newcommand{\cS}{\mathcal{S}}
\newcommand{\cT}{\mathcal{T}}

\newcommand{\cV}{\mathcal{V}}
\newcommand{\cW}{\mathcal{W}}
\newcommand{\cX}{\mathcal{X}}

\newcommand{\cZ}{\mathcal{Z}}

\newcommand{\ra}{\rightarrow}

\newcommand{\lB}{\left [}
\newcommand{\rB}{\right ]}
\newcommand{\lb}{\left (}
\newcommand{\rb}{\right )}

\newcommand{\where}{\text{where}}

\newcommand{\tand}{\text{and}}

\newcommand{\bensp}{\begin{eqnarray}\begin{split}}
\newcommand{\eensp}{\end{eqnarray}\end{split}}

\newcommand{\bnm}{\begin{matrix}}
\newcommand{\enm}{\end{matrix}}

\def\Xint#1{\mathchoice
{\XXint\displaystyle\textstyle{#1}}%
{\XXint\textstyle\scriptstyle{#1}}%
{\XXint\scriptstyle\scriptscriptstyle{#1}}%
{\XXint\scriptscriptstyle\scriptscriptstyle{#1}}%
\!\int}
\def\XXint#1#2#3{{\setbox0=\hbox{$#1{#2#3}{\int}$ }
\vcenter{\hbox{$#2#3$ }}\kern-.6\wd0}}

\def\dashint{\Xint-}

\newcommand{\ket}[1]{|#1\big>}
\newcommand{\bra}[1]{\big<#1|}

\newcommand{\braket}[2]{\big<#1|#2\big>}

\newcommand{\cs}{Chern-Simons\ }

\usepackage{wrapfig}

\begin{document}

\parindent=12pt

\begin{center}
    {\Large \bf Large \texorpdfstring{$N$}{N} Invariants of Torus Links in Lens Spaces}
\end{center}

\baselineskip=18pt

\bigskip

\centerline{Kushal Chakraborty and Suvankar Dutta}

\bigskip

\centerline{\large \it Indian Institute of Science Education and Research
Bhopal}
\centerline{\large \it Bhopal bypass, Bhopal 462066, India}

\bigskip

\centerline{E-mail: kushal16@iiserb.ac.in, suvankar@iiserb.ac.in}

\vskip .6cm
\medskip

\vspace*{4.0ex}

\centerline{\bf Abstract} \bigskip

\noindent 
We compute the invariants for a class of knots and links in arbitrary representations in $S^3/\mathbb{Z}_p$ in the large $k$ (level), large $N$ (rank) limit, keeping $N/(k+N)=\lambda$ fixed, in $U(N)$ and $Sp(N)$ Chern-Simons theories. Using the relation between the saddle point description and collective field theory, we first find that the invariants for the Hopf link and unknot are given by the on shell collective field theory action. We next show that the results of these two invariants can be used to compute the
invariants of other torus knots and links. We also discuss the large $N$ phase structure of the Hopf link invariant and observe that the same may admit a Douglas-Kazakov type phase transition depending on the choice of representations and $\lambda$.

\vfill\eject

\tableofcontents


\section{Introduction}
\label{sec:intro}
Knot theory is an interesting field in mathematics. A knot is a smooth embedding of a circle $\mathcal{S}^{1}$ to a three manifold . A link is a collection of disjoint knot.
An important question in knot theory is whether two knots (or links in general) are equivalent by ambient isotopy (i.e. if one can continuously deform one knot into the other without breaking it).
\emph{Knot invariants}\footnote{Knot invariants in general mean invariants for both knots and links.} are used to distinguish between different in-equivalent knots

A variety of knot invariants in three dimensions can be obtained from the correlation functions of Wilson loop (WL) operators in Chern-Simons (CS) theory \cite{Wittenjones}. The CS theory is topological at classical as well as quantum level. The only interesting observables in this theory are WLs along oriented knots and links. Since WLs are independent of metric, their expectation values are topological and hence they are bona fide candidates for the  knot invariants. Witten proved that these topologically invariant correlation functions are precisely the generalised knot invariants in three manifolds \cite{Wittenjones}. For $G\equiv SU(N)$, $G\equiv SU(2)$ and $G\equiv SO(N)$ these correlation functions give the HOMFLY-PT polynomial, the Jones polynomial and the Kauffman polynomials respectively when the representations associated with the WLs are in fundamental representations. In general the correlation functions of WLs in arbitrary representations provide a wide class of topological invariants. These are known as \emph{coloured polynomials}. From the CS theory perspective such invariants are well defined and we also understand the source of representations associated with the invariants. In mathematics such colored invariants have been defined by Reshetikhin and Turaev \cite{Reshetikhin} and used to construct new quantum invariants of 3-manifolds.

Although CS theory can be solved exactly, the calculation of knot invariants is difficult. 
\begin{wrapfigure}{r}{0.45\textwidth}
  \begin{center}
    \includegraphics[width=0.4\textwidth]{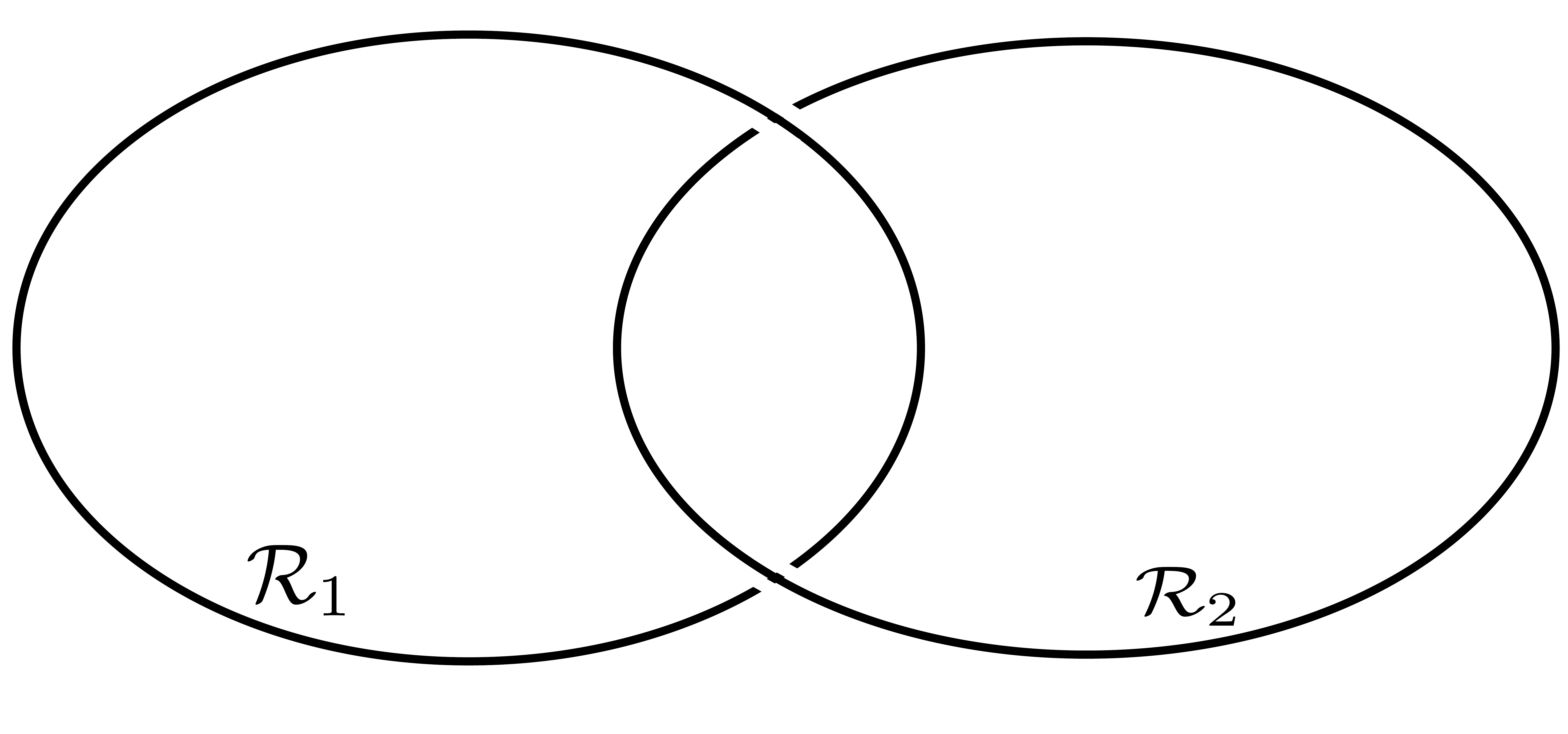}
  \end{center}
  \caption{Hopf link.}
  \label{fig:hopf}
\end{wrapfigure}
There exists a large literature on the computation of generalised coloured invariants for different links and knots. Some of them can be found in \cite{RamaDevi:1992np,Nawata:2013qpa,Ramadevi:1994zb,Ramadevi:1993hu,Labastida:1990bt,Labastida:2000zp,Labastida:2000yw,Isidro:1992fz,Singh:2021ehs,Mironov:2015era,Mironov:2016izi,Mironov:2014zza,Mironov:2015ota,Nawata:2013mzx,Kameyama:2019mdf,Hitoshi1}. In this paper we consider $U(N)$ and $Sp(N)$ CS theories and compute the knot invariants in the following limit 
\ben\label{eq:doublescaling}
N,k \ra \infty \quad \text{keeping $\lambda = \frac{N}{k+N}$ fixed}.
\een
Here, $N$ is the rank of the gauge group and $k$ is the level of the CS theory. This limit is called \emph{double scaling limit}. We use the saddle point technique to calculate the invariant for the Hopf link, shown in fig.\ref{fig:hopf} in lens space $S^3/\mathbb{Z}_p$ for any representations $\cR_1$ and $\cR_2$ associated with two unknots\footnote{The computation presented in this paper could be extended to find the invariants for hyperbolic knots and links. For example Borromean links (for 3 given representations $\cR_1$, $\cR_2$ and $\cR_3$).  Therefore such calculations might be helpful in understanding the volume conjecture \cite{Kashaev,Murakami,Hitoshi1,Gukov:2003na} and it's generalised version.}.
We then show that with this result in hand one can compute the invariants for the class of torus knots (2,$k$=odd) and links (2,$k$=even) following the method developed in \cite{RamaDevi:1992np}.

Since the representations $\cR_{1/2}$ are integrable, in the large $N$ limit we describe these representations in terms of eigenvalue distributions of unitary matrices. As a result, in the double scaling limit the problem can be studied with the aid of a collective field theory\footnote{Similar analysis was done by Gross and Matytsin \cite{Grossmatytsin,Matytsin} in the context of 2d Yang-Mills theory.} \cite{jevicki}. We find that the Hopf link invariant satisfies the Hamilton-Jacobi equation with $p \lambda$ playing the role of time and the Hamiltonian being the free collective field theory Hamiltonian. Therefore, large $N$ (classical) value of the Hopf link invariant, evaluated on the classical solutions, is given by the on shell action of the free collective field theory. The Hamilton's equations are similar to the continuity and Navier-Stokes equations of an incompressible fluid with negative pressure. Thus the study of Hopf link invariant boils down to the study of time evolution of an one dimensional incompressible fluid with the boundary conditions depending on $\cR_{1/2}$.

The knowledge of the Hopf link invariant should be enough to find the invariants for the class of other torus knots (2,$k$=odd) and links (2,$k$=even) using the generalised Alexander-Conway skein relations recursively.
 However when we place representations other than fundamental it becomes difficult. In \cite{RamaDevi:1992np} Ramadevi et.al. developed a technique to get the invariants for links made of braids up to four strands. A class of invariants for links and knots including torus knots can be obtained from the eigenvalues of half-twist braid matrices. We use these relations and show that in the double scaling limit the invariants for a wide class of knots and links can be computed from the result of  the Hopf link invariant.  We also emphasise that our results are valid for any arbitrary large $N$ representations associated with the knots.
 
We also study the large $N$ phase structure of two point correlation functions of CS theory. We show that in the double scaling limit the two point correlation function (Hopf link invariant) admits a third order phase transition depending on the choice of $\cR_{1/2}$ and $\lambda$. Such a phase transition is similar to the Douglas-Kazakov phase transition \cite{douglas-kazakov}.

\paragraph{Main results :}The main results of this paper are following.
\begin{itemize}
	\item We show that in the $N$ limit how the computation of two point correlation functions of CS theory for Hopf link boils down to computation of  on shell action of a free collective field theory. To be precise, the calculation of Hopf link invariant turns out to be equivalent to the time evolution of an one dimensional incompressible fluid with the boundary conditions depending on $\cR_{1/2}$.
	\item The Hopf link invariant in the large $N$ limit is equal to the on-shell partition function of a collective field theory. The result is given by eqn (\ref{eq:WR1R2final}). The functions $\sigma_{1/2}(\theta)$ contain information of the representations $\cR_{1/2}$ associated with the Hopf link.
	\item We also show that in the large $N$ limit the invariants for torus knots (2, $k$=odd) and links (2, $k$=even) can be computed from the Hopf link invariants and are given by (\ref{otherknot}).
	\item We study the large $N$ phase transition in two point correlation functions of CS theory.
	\\
\end{itemize}

The plan of the paper is as follows. In section \ref{sec:UNCScorrelators} we review the correlation functions of WLs in CS theory in Seifert manifold in different framings. Section \ref{sec:UNCS} provides the detailed calculation of the Hopf link invariant and other torus knot invariants in $U(N)$ CS theory. We also discuss the large $N$ phase structure of the Hopf link invariant in this section. We conclude the section \ref{sec:UNCS} with an example. Large $N$ phase structure of $Sp(N)$ CS theory and correlations functions are discussed in section \ref{sec:spn}. We end the paper with a discussion \ref{sec:disc}. In appendix \ref{app:surgery} we discuss how to obtain the correlation functions in different manifolds using surgery and their framing dependence. In other appendices we elaborate various technical details used in the main text.

\section{Preliminary : correlation functions in Chern-Simons theory}
\label{sec:UNCScorrelators}

In this section we review the correlation functions in CS theory on Seifert manifolds in different framings. Experts may skip this section.

The topological nature of classical CS theory is preserved even at the quantum level (correlation functions) but at the cost of a choice of \emph{framing}. Physical observables (correlations of WLs) are completely determined in terms of topological data of the three manifold $\mathbb{M}$ up to a framing. In order to understand the framing dependence in detail we first state an important connection between CS theory and Wess-Zumino-Witten (WZW) model \cite{Wittenjones}. Quantisation of CS theory with gauge group $G$ and level $k$ in $\mathbb{M}$ with boundary $\Sigma$ ($= \partial \mathbb{M}$) renders a physical Hilbert space $\cH(\Sigma)$ which is isomorphic to space of conformal blocks of WZW model with an affine Lie algebra $\mathbf{g}_k$. Using this connection, correlations of WL operators in CS theory can be written in terms of observables of the WZW model.

To show the framing dependence explicitly we consider the CS theory on a Seifert manifold $\cM_{(g,p)}$. A Seifert manifold is a non-trivial circle bundle over genus $g$ Riemann surface $\Sigma_g$ with the first Chern class $p$. Seifert manifolds for a generic $p$ can be obtained from $\cM_{(g,0)}$ (which is a product of genus $g$ Riemann surface and a circle $\Sigma_g \times S^1$) by surgery. Different choices of surgeries give different framings of $\cM_{(g,p)}$. The $n$-point correlation functions in CS theory are defined as
\ben\label{eq:ncorrel}
\langle \cW_{\cR_1,\cdots \cR_n}^{\mathrm{K}_1,\cdots, \mathrm{K}_n}\rangle & = \int [DA] e^{iS_{CS}} \prod_{a=1}^{n} \cW_{\cR_a}^{\mathrm{K}_a}(A)
\een
$\where \ \mathcal{W}_\cR^\mathrm{K}(A)=\Tr_\cR U_\mathrm{K} \ \text{with}\ U_\mathrm{K}=P \exp \oint_\mathrm{K} A$ and $S_{CS}$ is the standard CS action with gauge field $A$. Denoting the $n$-point correlations of WLs by $\cW^{(g,p)}_{\cR_1\cdots \cR_n}[G,k]$ one can write them in the following form \cite{Naculich:2007nc,blauthompson,Blau:2006gh}
\ben\label{eq:ncorrelWZW}
	\cW^{(g,p)}_{\cR_1\cdots \cR_n}[G,k] & = \sum_{\cR} \cK^{(p)}_{\cR_1\cR} \cW_{\cR\cR_2\cdots \cR_n}(\Sigma_g\times S^1,G,k)
\een
where $\cK^{(p)}$ depends on the framing. Note that being topological $\cW^{(g,p)}_{\cR_1\cdots \cR_n}[G,k]$ does not explicitly depend on the geometry of the knots $\mathrm{K_i}$s. It depends on the \emph{linking number} of the link made out of these knots. $\cW_{\cR\cR_2\cdots \cR_n}(\Sigma_g\times S^1,G,k)$ is the $n$-point correlation function in $\Sigma_g\times S^1$, given by
\ben\label{eq:s2s1correl}
\cW_{\cR\cR_2\cdots \cR_n}(\Sigma_g\times S^1,G,k) = \sum_{\cR} \cS_{0\cR}^{2-n-2g} \prod_{a=1}^n \cS_{\cR\cR_a}.
\een
See appendix \ref{app:surgery} for a detailed discussion. Here $\cS_{\cR\cR'}$ and $\cT_{\cR\cR'}$ are the modular transform matrices associated with the highest weight representations of the affine Lie algebra $\mathbf{g}_k$ under inversion and translation of modular parameter respectively (see \cite{yellowbook} for details). 
\begin{figure}[h]
  \begin{center}
    \includegraphics[width=0.4\textwidth]{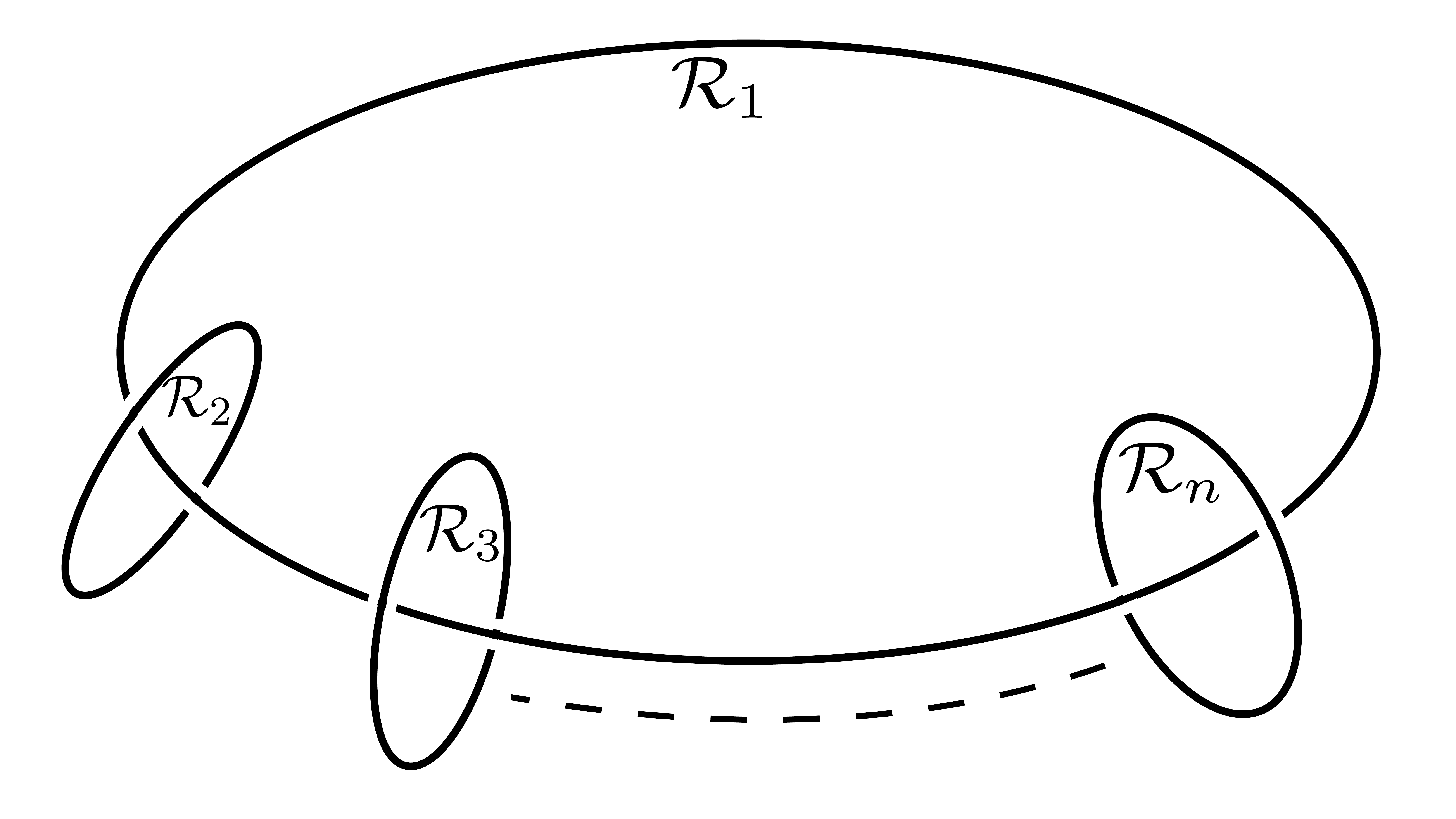}
  \end{center}
  \caption{Links of unknots.
  \, }
  \label{fig:nWL}
\end{figure}
They satisfy
\ben
\cS^2 = (\cS\cT)^3= \mathbb{I}.
\een
The summation on the right hand side of (\ref{eq:ncorrelWZW}) runs over integrable representations of $\mathbf{g}_k$. $\cW^{(g,p)}_{\cR_1\cdots \cR_n}[G,k]$ in (\ref{eq:ncorrelWZW}) represents a topological invariant of links of $n$ unknots as shown\footnote{The figure \ref{fig:nWL} describes the linking of the knots schematically. The exact diagrams of such links in $\cM_{(g,p)}$ is complicated. Mathematically the $p$ dependence in the correlation function comes (after surgery) through the operator $\cK^{p}$.} in fig \ref{fig:nWL}.

When $p=1$ and $g=0$ the Seifert manifold is a three sphere $S^3$. On $S^3$, there exists a canonical choice of framing, given by $\cK^{(1)}=\cS$. In canonical framing the correlation of $n$ WLs (\ref{eq:ncorrelWZW}) is given by
\ben
\cW_{\cR_1\cdots \cR_n}[S^3,G,k] = \cS^{2-n}_{0\cR_1} \prod_{a=2}^n \cS_{\cR_1\cR_a}.
\een

The invariant for the link diagram in fig \ref{fig:nWL} in Seifert manifold can be obtained from $\Sigma_g\times S^1$ by surgery with the choice $\cK^{(p)}= (TST)^p$. This particular choice is called Seifert framing. In Seifert framing, therefore, the invariant is given by \cite{Naculich:2007nc,blauthompson,Blau:2006gh}
\ben\label{eq:WLonSM}
\cW^{(g,p)}_{\cR_1\cdots \cR_n}[G,k] = \sum_{\cR} \cT_{\cR \cR}^{-p} \cS_{0\cR}^{2-n-2g} \prod_{a=1}^{n} \cS_{\cR\cR_a}.
\een
The main focus of this paper is to study the large $N$ structure of these knot invariants.

\section{Knot invariants in \texorpdfstring{$U(N)$}{UN} Chern-Simons theory}\label{sec:UNCS}

In this section we consider the gauge group $G=U(N)$. For our purpose, we define a modified WL operator $\widetilde W_\cR$
\be\label{eq:modifiedWL}
\widetilde \cW_\cR = \frac{\cW_\cR}{\cS_{0\cR}}.
\ee
The $n$-point correlation functions of these modified WL operators are given by,
\ben
\widetilde \cW^{(g,p)}_{\cR_1 \cdot\cdot \cR_n}(N,k) = \sum_{\cR} \cS_{0\cR}^{2-2g-n} T_{\cR\cR}^{-p} \prod_{a=1}^n \frac{\cS_{\cR \cR_a}}{\cS_{0\cR_a}}.
\een
$\cS$ and $\cT$ for $u(N)_k$ are given by (\ref{eq:Smod}, \ref{eq:Trr}). For a given representation $\cR_a$ we define a set of $N$ variables $\{\theta^{(a)}_1,\cdots \theta^{(a)}_N\}$ 
\begin{equation}\label{eq:thetaidef}
    \theta_{i}^{(a)} = \frac{2\pi}{N+K}\lb h_{i}^{(a)}-\frac{N-1}{2}\rb 
\end{equation}
where $h_i^{(a)} = n_i^{(a)}+N-i$ are the hook numbers of the Young diagram associated with $\cR_a$ and $n_i^{(a)}$ is number of boxes in the $i^{th}$ row. Since $\cR_a$s are integrable representations of $u(N)_k$, $n_i^{(a)}$s satisfy
\ben\label{eq:integrabilitycond}
-\frac{k}{2} \leq n_N^{(a)}\leq \cdots \leq n_1^{(a)}\leq \frac{k}{2}.
\een
It is easy to check that in the double scaling limit the variables $\theta_i^{(a)}$ ranges from $-\pi$ to $\pi$. These $\theta^{(a)}_i$ can be thought of as eigenvalues of unitary matrices. However, there is a difference. The minimum gap between two eigenvalues is $2\pi/(k+N)$. Therefore, in large $N$ limit an integrable representation can be described in terms of a distribution of these eigenvalues, denoted by $\sigma_{(a)}(\theta)$ and defined as 
\ben\label{eq:sigmadef}
\sigma_a(\theta) = \frac1N \sum_{i=1}^N \delta (\theta - \theta_i^{(a)}) .
\een
In the double scaling limit (\ref{eq:doublescaling}), $\sigma_a(\theta)$ satisfies the constraint
\ben\label{eq:capcondition}
\sigma_a(\theta) \leq \frac{1}{2\pi \lambda}
\een
which follows from the fact that two eigenvalues have a minimum separation $2\pi/(k+N)$. Considering the above change of variables one can compute the ratio of $\cS_{\cR\cR_a}$ and $\cS_{0\cR_a}$. It turns out to be
\ben
\frac{\cS_{\cR\cR_a}}{\cS_{0\cR_a}} = e^{-\frac{2i}{N}\sum_i (h_i-\frac{N-1}{2}) \sum_i \theta^{(a)}_i}\chi_{R}(\theta^{(a)}_i)
\een
where $h_i$s are the hook numbers associated with $\cR$ and $\chi_\cR(\theta)$ is the character of the $U(N)$ in the representation $\cR$. Using the expression for $\cT$ (\ref{eq:Trr}) the $n$-point correlation function can be written as,
\ben\label{eq:WLntilde}
\widetilde \cW^{(g,p)}_{\cR_1 \cdot\cdot \cR_n}(N,k)= \sum_{\cR} \frac{q^{-\frac{p}{2} \cC_2(\cR)} e^{-i \Theta \cC_1(\cR)}}{\cS_{0\cR}^{2g+n-2}} \prod_{a=1}^n \chi_\cR(\theta^{(a)}).
\een
where
\ben\label{eq:qdef}
q=e^{2\pi i/(k+N)}
\een
and 
\ben
\Theta = 2 \sum_{a=1}^n \lb \frac{1}{N}\sum_{i=1}^{N}\theta^{(a)}_i \rb.
\een
This correlation function is similar to the partition function of a \emph{q}-deformed Yang-Mills theory with a $\Theta$-term on a genus $g$ Riemann surface with $n$ boundaries \cite{Grossmatytsin,Naculich:2007nc}. The distributions of holonomies on those boundaries are given by $\{\theta^{(a)}_i\}$. The only difference between (\ref{eq:WLntilde}) and the partition function of a \emph{q}-deformed 2d Yang-Mills theory is that the sum in (\ref{eq:WLntilde}) runs over integrable representations.

When $\cR_1= \cdots = \cR_n =0$, the correlation function (\ref{eq:WLntilde}) gives the partition function of $U(N)$ CS theory of level $k$ on Seifert manifold. The large $N$ phase structure of this theory for $g=0$ was studied in \cite{Chakraborty:2021oyq}. The theory undergoes a third order phase transition in the double scaling limit (\ref{eq:doublescaling}). We have reviewed the result in appendix \ref{app:pfreview}.

In this section we explicitly compute the two point correlation functions ($n=2$) in the \emph{double scaling limit} (\ref{eq:doublescaling}) and show that
 using the result of two point correlation function one can compute the invariants for a class of torus knots (2,k=odd) and links (2,k=even).

Two point correlators in $S^3/\mathbb{Z}_p$, which is a Seifert manifold with $g=0$ for any $p$, is given by
\ben\label{eq:W2}
\cW_{\cR_1,\cR_2}[S^3/\mathbb{Z}_p,N,k] = \sum_{\cR} \cT_{\cR \cR}^{-p} \cS_{\cR\cR_1} \cS_{\cR\cR_2} 
\een
and represents a topological invariant for the Hopf link (fig \ref{fig:hopf}). In the large $N$ limit the two point function admits a genus expansion (perturbative part) \cite{Mironov:2013oma}
\ben\label{eq:genexp}
\ln\lB\cW_{\cR_1,\cR_2}[S^3/\mathbb{Z}_p,N,k]\rB = \sum_{h=0}^\infty N^{2-2h} \mathrm{W}^{(h)}_{\cR_1,\cR_2}(\lambda).
\een
Our goal is to compute the leading contribution  $\mathrm{W}^{(0)}_{\cR_1,\cR_2}(\lambda)$ for any $\cR_{1/2}$ using the saddle point technique. 

The modified two point correlation function (\ref{eq:modifiedWL}) is given by
\begin{equation}\label{eq:tildeWL2}
    \begin{split}
     \widetilde \cW_{\cR_1\cR_2}(S^3/\mathbb{Z}_p,N,k)
      = \sum_{\cR}  q^{-\frac{p}{2} \cC_2(\cR)} e^{-i \Theta \cC_1(\cR)} \chi_\cR(\theta^{(1)}) \chi_\cR(\theta^{(2)})
    \end{split}
\end{equation}
where $\theta^{(1/2)}$ are eigenvalues corresponding to the representation $\cR_{1/2}$. We consider the representations $\cR_{1}$ and $\cR_{2}$ such that the eigenvalues are symmetrically distributed about \emph{zero} and hence the $\Theta$ term drops out from (\ref{eq:tildeWL2}). Expressing the characters of $U(N)$ in terms of Schur polynomial we can explicitly write down the two point correlation functions in the following form,
\begin{equation}\label{eq:cor}
 \begin{split}
     \widetilde \cW_{\cR_1\cR_2}( S^3/\mathbb{Z}_p,N,k)
     = \sum_{\{y_{i}\}}\frac{e^{2\pi i p\frac{N^{2}}{24}}}{2^{N(N-1)}}\frac{\det||e^{i N y_{j}\theta^{(1)}_{k}}||}{\prod_{i<j}^{N}\sin\left(\frac{\theta^{(1)}_{i}-\theta^{(1)}_{j}}{2}\right)} \frac{\det||e^{i N y_{j}\theta^{(2)}_{k}}||}{\prod_{i<j}^{N}\sin\left(\frac{\theta^{(2)}_{i}-\theta^{(2)}_{j}}{2}\right)}
   \  e^{-p\pi i N\lambda\sum_{j=1}^{N}y_{j}^{2}}
     \end{split}
 \end{equation}
where 
\ben\label{eq:yihirel}
y_{j}=\frac{1}{N}\left(h_{j}-\frac{N-1}{2}\right)
\een
is the shifted hook number for the representation $\cR$.

In general it is difficult to find an exact expression for the two point function. However, in the large $N$ limit we see that the right hand side of (\ref{eq:cor}) is dominated by a single representation for a specific class of $\cR_1$ and $\cR_2$ and it is indeed possible to find an exact expression for $\widetilde \cW_{\cR_1\cR_2}( S^3/\mathbb{Z}_p,N,k)$.

\subsection{Large \texorpdfstring{$N$}{N} analysis of two point function : invariant for the Hopf link}
\label{sec:largeN}

We follow the work of Gross and Matytsin \cite{Grossmatytsin,Matytsin} to do the large $N$ analysis of (\ref{eq:cor}). In the large $N$ limit the integrable representation on the right hand side of (\ref{eq:tildeWL2}) can be denoted by a density function $\rho(y)$, 
\ben
\rho(y) = - \frac{dx}{dy(x)}, \quad \text{where} \quad y(x)=\frac{y_i}{N}, \ x=\frac{i}{N}.
\een
Since $i$ runs from 1 to $N$, $x\in [0,1]$. Also from the discreteness of $y_i$s (eqn. \ref{eq:yihirel}) it follows that
\ben
\rho(y) \leq 1.
\een
Following \cite{Chakraborty:2021oyq}, after a wick rotation in complex $p$ plane $(p\to - i p)$ we introduce\footnote{Since the correlation function is an analytic function of $p$ one can Wick rotate the final answer back to the original $p$.}
\ben\label{eq:Zdef}
\widetilde \cZ_{\cR_1\cR_2} &=& 2^{N(N-1)}e^{-\pi p N^2/12} \widetilde \cW_{\cR_1\cR_2}( S^3/\mathbb{Z}_p,N,k)
\een
and a new variable $A$ in place of $p \lambda$
\ben\label{eq:Adef}
A = 2\pi p\lambda .
\een
One can show from (\ref{eq:cor}) that $\widetilde \cZ_{\cR_1\cR_2}$ satisfies
\begin{equation}\label{eq:den}
    \begin{split}
        2N\frac{\partial \widetilde{\cZ}_{\cR_{1}\cR_{2}}}{\partial A}=\frac{1}{D[\theta^{(a)}]}\sum_{k=1}^{N}\frac{\partial^{2}}{\partial\theta_{k}^{(a)^{2}}}[D[\theta^{(a)}]\widetilde{\cZ}_{\cR_{1}\cR_{2}}]
    \end{split}
\end{equation}
where
\begin{equation}
   D[\theta^{(a)}]= \prod_{i<j}^{N}\sin\left(\frac{\theta^{(a)}_{i}-\theta^{(a)}_{j}}{2}\right) \quad \text{for $a=1,2$}.
\end{equation}
Assuming $\widetilde \cZ_{\cR_1\cR_2}$ is dominated by a single representation in the large $N$ limit, we choose
\ben\label{eq:FNdef}
\widetilde{\cZ}_{\cR_{1}\cR_{2}}=e^{N^{2}F_{N}} \quad \text{and $\lim_{N\to\infty}F_{N}=F$}
\een
and after a little algebra we find that $F$ satisfies (see appendix \ref{app:Fcalculation} for detailed calculation)
\begin{equation}\label{eq:free1}
\begin{split}
2\frac{\partial F}{\partial A} = \int\sigma_{a}(\theta)\Big[\Big(\frac{\partial }{\partial\theta}\frac{\delta F}{\delta\sigma_{a}(\theta)}\Big)+2U(\theta)\frac{\partial}{\partial\theta}\frac{\delta F}{\delta\sigma_{a}(\theta)}
+U(\theta)^{2}-\frac{\pi^{2}}{3}\sigma_{a}(\theta)^{2}\Big].
\end{split}
\end{equation}
Considering $F$ to be of the following form\footnote{Here the cut-integral $\Xint - d\theta'$ represents integration over $\theta'$ over the valid range except the point $\theta$. Such integrals sometimes are denoted by $P \int$.} 
\begin{equation}\label{eq:free}
            F[\sigma_{1}(\theta), \sigma_{2}(\theta)|A] 
        = S[\sigma_{1}(\theta), \sigma_{2}(\theta)|A]
        -\frac{1}{2} \sum_{a=1}^2 \int\sigma_{a}(\theta)d\theta\dashint\sigma_{a}(\theta') \log[\sin\Big(\frac{\theta-\theta'}{2}\Big)]d\theta'
 \end{equation}
one can recast the equation (\ref{eq:free1}) as
\begin{equation}{\label{eq:hj}}
\begin{split}
   \frac{\partial S}{\partial A}=\frac{1}{2} \int\sigma_{a}(\theta)\Big[\Big(\frac{\partial}{\partial\theta}\frac{\delta S}{\delta\sigma_{a}(\theta)}\Big)^{2}-\frac{\pi^{2}}{3}\sigma_{a}(\theta)^{2}\Big]d\theta .
\end{split}
\end{equation}
As discussed in \cite{Grossmatytsin} this equation can be thought of as the Hamilton–Jacobi equation with Hamiltonian
\begin{equation}\label{eq:Ham}
    \begin{split}
        H[\sigma,\Pi]=\frac{1}{2}\int\sigma(t,\theta)\Big[\Big(\frac{\partial\Pi(t,\theta)}{\partial\theta}\Big)^{2}-\frac{\pi^{2}}{3}\sigma(t,\theta)^{2}\Big]d\theta
    \end{split}
\end{equation}
of a $(1+1)$ dimensional field theory with field $\sigma(t,\theta)$ and conjugate momentum $\Pi(t, \theta)=\frac{\delta S}{\delta\sigma(t, \theta)}$.
The Hamilton's equations of motion are given by,
\begin{equation}\label{eq:eu}
     \begin{split}
         \frac{\partial \sigma}{\partial t}+\frac{\partial (\sigma v)}{\partial\theta} & =0\\
         \frac{\partial v}{\partial t}+v\frac{\partial v}{\partial \theta} - \pi^{2}\sigma\frac{\partial\sigma}{\partial\theta} & = 0
     \end{split}
 \end{equation}
where
\ben
v(t,\theta) = \frac{\partial \Pi(t,\theta)}{\partial \theta}.
\een
These equations are similar to the continuity and Navier-Stokes equations of an one dimensional fluid moving on a circle with density $\sigma(t,\theta)$, velocity $v(t,\theta)$ and a negative pressure. Finding a solution of the equation (\ref{eq:hj}) for $S(\sigma_1,\sigma_2)$ is equivalent to solve these fluid equations with the boundary conditions 
\begin{equation}\label{eq:boundarycond}
   \begin{split}
       \sigma(t=0,\theta) & = \sigma_{1}(\theta)\\
       \sigma(t=A, \theta) & = \sigma_{2}(\theta).
   \end{split}  
 \end{equation}
Suppose $(\bar\sigma(t,\theta),\bar v(t,\theta))$ is a solution of fluid equations with the desired boundary conditions. The quantity $S(\sigma_1,\sigma_2,A)$ evaluated on this solution is therefore given by
\ben
\bar S(\sigma_1,\sigma_2,A) = \frac{1}{2}\int H[\bar\sigma(0,\theta), \bar v(0,\theta)] dA + \text{const}.
\een
Using the series of definitions (\ref{eq:free}), (\ref{eq:FNdef}) and (\ref{eq:Zdef}) one can write the knot invariant for Hopf link in $S^3/\mathbb{Z}_p$ as
\ben
\begin{split}
\cW_{\cR_1\cR_2}  = \cS_{0\cR_1}\cS_{0\cR_2} \widetilde\cW_{\cR_1\cR_2} 
 = \exp\lB N^2 \lb \bar S(\sigma_1,\sigma_2,A) + \frac{\pi p}{12} -\ln 2 \rb \rB.
\end{split}
\een
Hence from (\ref{eq:genexp}) we find
\ben\label{eq:WR1R2final}
\mathrm{W}^{(0)}_{\cR_1\cR_2} = \bar S(\sigma_1,\sigma_2,A) + \frac{\pi p}{12} -\ln 2 .
\een
Thus we see that the invariant for the Hopf link with any arbitrary large representations of $u(N)_k$ is given by the on shell free collective field theory action with the boundary conditions (\ref{eq:boundarycond}). Further, since the fluid equations (\ref{eq:eu}) are dispersionless KdV equation (Burger equation), the whole exercise to find the Hopf link invariant boils down to solving a dispersionless KdV equation with a set of boundary conditions. According to our notation if  $\cR_2$ (or $\cR_1$) is zero, then the corresponding two point function $\cW_{\cR_1\cR_2}$ becomes one point function and gives the invariant for unknot. $\cR_2=0$ corresponds to a $\sigma_2$ given by (\ref{eq:sigmanobox}).

Our next goal is to find out the invariants for other class of torus knots with the result of the Hopf link invariant in our disposal.

\subsection{Invariants for a class of torus knots in \texorpdfstring{$S^3$}{S3}}
\label{sec:torusknot} 

In this section we discuss how knot invariants for other torus knots in $S^3$ (a Seifert manifold with $g=0$, $p=1$) can be computed in the large $N$ limit from the result of the Hopf link.

\begin{figure}
     \centering
     \begin{subfigure}{0.47\textwidth}
         \centering
         \includegraphics[scale=.35]{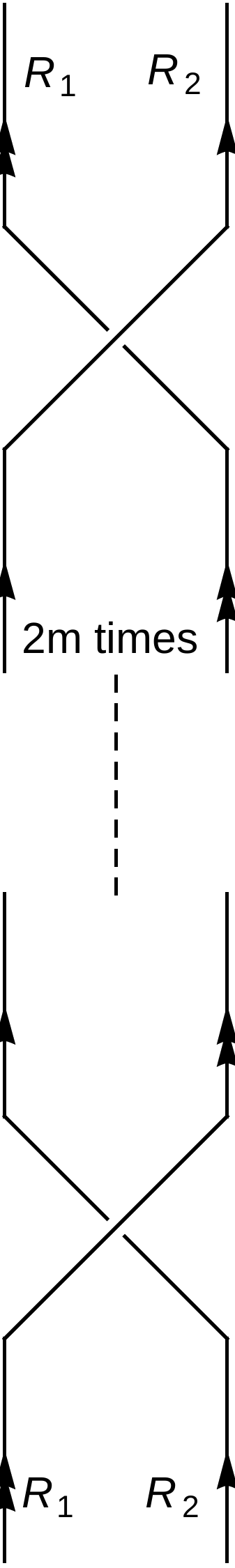}
         \caption{For any values of $m$ such braiding gives $m$ links of two unknots in representations $\cR_1$ and $\cR_2$ with the same orientations.}
         \label{fig:L2m}
     \end{subfigure}
     \hfill
          \begin{subfigure}{0.47\textwidth}
         \centering
         \includegraphics[scale=.35]{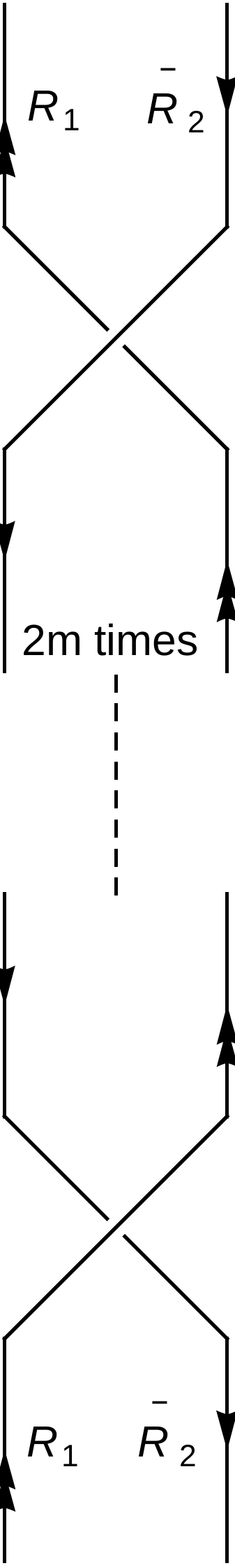}
         \caption{For any values of $m$ such braiding gives $2m$ crossings of two unknots in representations $\cR_1$ and $\cR_2$ with opposite orientations.}
         \label{fig:L2mb}
     \end{subfigure}
        \caption{Braiding of knots in $S^3$}
        \label{fig:three graphs}
\end{figure}

Torus knots are special kinds of knots which can be put on the surface of a torus. A formalism of knot operators was developed in \cite{Labastida:1990bt} to compute the invariants of torus knots. In \cite{RamaDevi:1992np} the authors developed a different method to obtain the invariants of links made from braids of up to four strands. It was shown in \cite{RamaDevi:1992np} that for special types of links as shown in figure \ref{fig:L2m} and \ref{fig:L2mb}, the knot invariants can be written in terms of the eigenvalues of the half-twist matrix.
The half-twist matrix $\cB(\cR_1,\cR_2)$ introduces right-handed half-twists in parallelly oriented strands carrying representations $\cR_1$ and $\cR_2$ whereas, $\hat \cB(\cR_1,\bar\cR_2)$ introduces right-handed half-twists in oppositely oriented strands carrying representations $\cR_1$ and $\bar \cR_2$. The dimensions of $\cB$ and $\hat\cB$ depend on the number of irreducible representations in the product $\cR_1$ and $\cR_2$ ($\bar \cR_2$). Denoting the link invariants in fig \ref{fig:L2m} and \ref{fig:L2mb} by $\cV[\cL_{2m}(\cR_1,\cR_2)]$ and $\cV[\hat\cL_{2m}(\cR_1,\bar\cR_2)]$ respectively it was shown in \cite{RamaDevi:1992np} that they are given by
\ben\label{eq:L2m}
\cV[\cL_{2m}(\cR_1,\cR_2)] = \sum_{\cR} \mathrm{dim}_q \cR \ (\lambda^+_\cR(\cR_1,\cR_2))^{2m}
\een
and
\ben\label{eq:L2mb}
\cV[\hat\cL_{2m}(\cR_1,\bar \cR_2)] = \sum_{\cR} \mathrm{dim}_q \cR \ (\lambda^-_\cR(\cR_1,\bar\cR_2))^{2m}
\een
where $\lambda^{\pm}_\cR(\cR_1,\cR_2)$ are the eigenvalues of $\cB(\cR_1,\cR_2)$ and $\hat \cB(\cR_1,\bar\cR_2)$ respectively. They are given by
\ben
\begin{split}
\lambda^+_\cR(\cR_1,\cR_2) & = (-1)^{\epsilon_\cR} q^{\cC_2(\cR_1)+\cC_2(\cR_2) + \frac{|\cC_2(\cR_1)-\cC_2(\cR_2)|}{2}- \frac{\cC_2(\cR)}{2}}\\
\lambda^-_\cR(\cR_1,\bar\cR_2) & = (-1)^{\epsilon_\cR} q^{ -\frac{|\cC_2(\cR_1)-\cC_2(\cR_2)|}{2}+ \frac{\cC_2(\cR)}{2}}
\end{split}
\een
where $q$ is given by (\ref{eq:qdef}). The sum on the right hand side runs over distinct irreducible representations of $\cR_1 \otimes \cR_2 (\bar \cR_2)$. The factor $(-1)^{\epsilon_\cR}$ depends on the symmetric or anti-symmetric properties of $\cR$ in the product of $\cR_1 \otimes \cR_2$. For $m=1$, $\cV[\cL_{2m}(\cR_1,\cR_2)]$ and $\cV[\hat\cL_{2m}(\cR_1,\cR_2)]$ give the invariants for the Hopf link in $S^3$ with the same and opposite orientations respectively. 

Denoting
\ben\label{eq:L2ma}
\cG(\cR_1,\cR_2) = q^{\cC_2(\cR_1)+\cC_2(\cR_2) + \frac{|\cC_2(\cR_1)-\cC_2(\cR_2)|}{2}}
\een
the knot invariant $\cV[\cL_{2m}(\cR_1,\cR_2)]$ can be written as
\ben
\cV[\cL_{2m}(\cR_1,\cR_2)] = \cG(m,\cR_1,\cR_2) \sum_{\cR} \mathrm{dim}_q \cR \ q^{- m \cC_2(\cR)}.
\een
In the double scaling limit (\ref{eq:doublescaling}), using the change of variables (\ref{eq:thetaidef}) and denoting the large representations $\cR_{1/2}$ by $\sigma_{1/2}$ respectively the invariants (\ref{eq:L2ma}) can be written as,
\ben
\cV[\cL_{2m}(\cR_1,\cR_2)] = \hat \cG(m,\lambda, \sigma_1,\sigma_2) \cF(m,\lambda)
\een
where
\begin{align}
  \hat \cG(m,\lambda, \sigma_1,\sigma_2) & = \exp \Bigg[ \frac{i  m N^2}{2\pi\lambda} \lb  \int d\theta \theta^2 \lb \sigma_1+\sigma_2 \rb + \frac12 \left| \int d\theta \theta^2  (\sigma_1-\sigma_2) \right|\rb  -\frac{i \pi m \lambda N^2}{3} \Bigg] \nonumber\\
\end{align}

and
\ben \label{eq:Ffunc}
\cF(m,\lambda) = \int \lB \cD\theta(x)\rB \exp \lB \frac{N^2}{2}\int dx \Xint - dy \log \sin \left| \frac{\theta(x)-\theta(y)}{2} \right |- \frac{i m N^2}{2\pi \lambda} \int dx \ \theta(x)^2\rB. \qquad
\een
The function $\cF(m,\lambda)$ is difficult to calculate as the functional integration over $\theta(x)$ does not run over all possible configurations (since $\cR$ in (\ref{eq:L2m}) runs over irreducible representations of $\cR_1 \otimes \cR_2 (\bar \cR_2)$ only). However in the large $N$ limit one can find the invariant for any $m$ knowing the value of the same for $m=1$. As we have mentioned that for $m=1$, $\cV[\cL_{2m}(\cR_1,\cR_2)]= \cW_{\cR_1\cR_2}(\lambda)$, the invariant for the Hopf link evaluated in the last section\footnote{Up to an analytic continuation in $\lambda : \lambda\ra i \lambda$. This is due to the fact that in deriving $\cW_{\cR_1\cR_2}$ we did an analytic continuation in $p$.}. Now from the expression of the function $\cF(m,\lambda)$ we see that
\ben
\cF(m,\lambda) = \cF(1,\lambda/m)
\een
since the space of functional integration over $\theta(x)$ remains unchanged (for given $\cR_1$ and $\cR_2$) as we vary $m$. Therefore we find that in the double scaling limit the link invariants for the class of links shown in figure \ref{fig:L2m} are given by
\ben\label{otherknot}
\cV[\cL_{2m}(\cR_1,\cR_2)] = \lb \frac{
\hat \cG(m,\lambda,\sigma_1,\sigma_2)}{\hat \cG(1,\lambda/m,\sigma_1,\sigma_2)}\rb \cW_{\cR_1\cR_2}(\lambda/m).
\een
\begin{wrapfigure}{r}{0.4\textwidth}
  \begin{center}
    \includegraphics[width=0.075\textwidth]{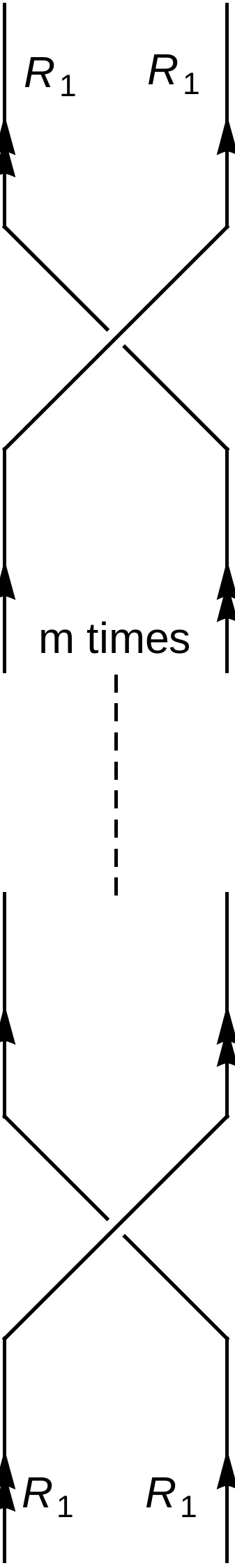}
  \end{center}
  \caption{For any odd values of $m$, such braiding gives a knot in representations $\cR_1$ with $m$ crossings.}
  \label{fig:Lm}
\end{wrapfigure}
In the similar way it is also possible to compute the link invariants $\cV[\hat\cL_{2m}(\cR_1,\bar\cR_2)]$ from $\cW_{\cR_1\bar\cR_2}(\lambda)$.

When $\cR_2=\cR_1$, one can construct a different braiding as shown in fig (\ref{fig:Lm}). For even $m$ this is same as braiding in fig (\ref{fig:L2m}) with $\cR_2=\cR_1$. However for odd values of $m$ such braiding gives a single knot. For example, when $m=3$ we get trefoil in $S^3$. $m=1$ gives an unknot. The invariant for this knot is given by
\ben\label{eq:Lm}
\cV[\cL_{m}(\cR_1,\cR_1)] = \sum_{\cR} \mathrm{dim}_q \cR \ (\lambda^+_\cR(\cR_1,\cR_1))^{m}\quad.
\een
Following the same argument, $\cV[\cL_{m}(\cR_1,\cR_1)]$ for any odd $m$ can be evaluated from $\cV[\cL_{1}(\cR_1,\cR_1)]$ in the double scaling limit. The knot invariant $\cV[\cL_{1}(\cR_1,\cR_1)]$ is given by (\ref{eq:WR1R2final}) with $\sigma_2$ representing a trivial representation (\ref{eq:sigmanobox}).\\

\subsection{Large \texorpdfstring{$N$}{N} phases of two point correlators in \texorpdfstring{$U(N)$}{UN} Chern-Simons theory}
\label{sec:domrep}

Large $N$ phase transition in $U(N)$ CS theory has been discussed in \cite{Chakraborty:2021oyq}. Similar phase transition is also observed in the correlation functions. In this section we give qualitative arguments for such phase transitions.

It can be shown that if the fluid equations (\ref{eq:eu}) admit a solution such that there exists a \emph{time} $0\leq t^*\leq A$ when the velocity of the fluid is zero, i.e.
\ben
v(t^*,\theta) =0
\een
then the fluid velocity and density satisfy an identity
\begin{equation}\label{eq:in}
i\pi\sigma^{*}[\theta-(t-t^*)(v(t,\theta)+i\pi\sigma(t,\theta))]=v(t,\theta)+i\pi\sigma(t,\theta))
\end{equation}
where
\begin{equation}
\sigma^{*}(\theta) = \sigma(t^*,\theta).
\end{equation}
The maximum value of the density is minimum at $t=t^*$. For such solution one can show that the two point function (\ref{eq:cor}) is dominated by an integrable representation $\rho(y)$ given by (see appendix \ref{app:evtoYD} for details)
\begin{equation}\label{eq:identity1}
\pi \rho(-\pi\sigma^{*}(\theta))=\theta 
\end{equation}
i.e. inverse of $\sigma^*(\theta)$. If the fluid equations do not admit any such solution then there exists no real saddle points. Needless to mention the existence of a real saddle point depends on the choice of $\sigma_{1/2}$ and $\lambda$.

If the fluid equations admit the existence of a real saddle point for a given $\sigma_{1/2}$ then depending on the value of $\lambda$ the system may undergo a phase transition. In order to discuss this phase transition qualitatively we assume that the initial and final densities $\sigma_1(\theta)$ and $\sigma_2(\theta)$ are even functions of $\theta$ and they have gaps (i.e. vanishes for $|\theta|$ greater than some $|\theta_0|\le \pi$). The fluid density $\sigma(t,\theta)$ starting from the configuration $\sigma_1(\theta)$ at $t=0$ spreads out (i.e. the gap starts decreasing). The absolute value of velocity of the fluid also decreases with time. At some intermediate time $t^*$ when the velocity of the fluid is zero the density has a maximum spread. After that fluid velocity starts increasing (in the opposite direction) and the density starts contracting and reaches the final configuration $\sigma_2(\theta)$ at $t=A$. The system will observe no phase transition if the maximum spread of $\sigma(t,\theta)$ at $t=t^*$ is less than $\pi$ or at max touches $\pi$. Otherwise the system will undergo a phase transition. The Young diagram density $\rho(y)$ and  have an upper cap $\rho(y) \le 1$. In addition the variable $y$ ranges between $\pm 1/2\lambda$. Since the two functions $\sigma^*(\theta)$ and $\rho(y)$ are functional inverse of each other, $\sigma^*$ having a gap means $\rho<1$. If $\sigma^*$ is gap-less then $\rho$ develops a cap.  Depending on the initial conditions and $\lambda$ we can have four possibilities.
\begin{itemize}
	\item If $\sigma^*(\theta)$ has gap but no cap then $\rho(y)$ has a no cap but a gap.
	\item If $\sigma^*(\theta)$ is gap-less with no cap then the dominant Young diagram $\rho(y)$ will have a cap but it is gap-less.
	\item If $\sigma^*(\theta)$ has a cap with a gap then $\rho(y)$ has no gap but no cap. 
	\item If $\sigma^*(\theta)$ has a cap and no-gap then $\rho(y)$ has a cap and no-gap.
\end{itemize}
The last two cases are special in CS theory (unlike 2d YM theory \cite{Grossmatytsin}) as all the representations are integrable representations. In the next section we elaborate this qualitative discussion with an example.

\subsection{An example : explicit computation of two point function and study of phase structure for Wigner semicircle distributions}
\label{sec:semicircle}

Wigner semicircle distribution is a probability distribution on real line between $\pm R$
\ben
f_R(x) = \frac{2}{\pi R^2}\Re \lb \sqrt{R^2-x^2}\rb.
\een
Such distributions appear in different contexts in physics and mathematics. The eigenvalue density of Gaussian hermitian random matrix theory is given by the semicircle distribution. The transition distribution of the limit shape of asymptotic Young diagrams studied by Vershik-Kerov and Logan-Shepp \cite{VerKer77,LogShe} is given by semicircle distribution. In this section we consider the large $N$ distribution $\sigma_{1/2}(\theta)$ corresponding to $\cR_1$ and $\cR_2$ are given by the semicircle distributions. The reason behind this choice is that the fluid equations can be solved exactly. Our goal is to explicitly calculate the knot invariant for the Hopf link for such representations in the large $N$ limit and study the phase transition of two point function.

We take the following semicircular distributions for the Young diagram density corresponds to $\cR_{1/2}$
\ben
\rho_{\cR_1/\cR_2} = \frac{1}{\pi}\sqrt{L_{1/2}-\frac{L_{1/2}^{2}y^{2}}{4}}.
\een
Since $\cR_{1/2}$ are integrable representations $L_{1/2}$ satisfies
\ben
L_1,L_2 \le \pi^{2}\quad \tand \quad L_1,L_2 \ge 16 \lambda^{2}.
\een
Suppose $u(y,t)$ denotes Young diagram distributions that interpolates between $\rho_{\cR_1}$ and $\rho_{\cR_2}$ from $t=0$ to $t=A$. We consider
\begin{equation} 
    u(y,t)=\frac{1}{\pi}\sqrt{\mu(t)-\frac{\mu(t)^{2}y^{2}}{4}}
\end{equation}
such that
\ben\label{eq:mubc}
\mu(0)=L_1 \quad \tand \quad \mu(A)=L_2.
\een
Considering the relation $\theta=2\pi \lambda y$, the eigenvalue distribution $\sigma(\theta,t)$ corresponding to $u(y,t)$ is given by 
\begin{equation}\label{eq:sigmaansatz}
    \sigma(\theta,t)=\frac{1}{2\pi \lambda}u\left(\frac{\theta}{2\pi\lambda},t\right) = \frac{1}{2\pi^2 \lambda} \sqrt{\mu(t) -\frac{\mu(t)^2 \theta^2}{16 \pi^2 \lambda^2}}.
\end{equation}

For the choice of fluid density (\ref{eq:sigmaansatz}) the fluid velocity can be solved exactly from the continuity equation and is given by
\ben\label{eq:vsol}
v(t,\theta) = - \theta \lb \frac{\dot \mu(t)}{2\mu(t)}\rb
\een
where $\dot \mu(t)$ is derivative of $\mu(t)$ with respect to $t$. Plugging the ansatz for $\sigma(t,\theta)$ and the solution for $v(t,\theta)$ in the Navier-Stokes equation we find a differential equation for $\mu(t)$
\ben\label{eq:mueom}
\Ddot{\mu}(t)-\frac{3 \dot\mu(t)^2}{2 \mu (t)}-\frac{1}{32\pi^{4}\lambda^{4}} \mu (t)^3 = 0.
\een
This is a second order non-linear ordinary differential equation and has a simple solution up to two integration constants. These two constants can be fixed from the boundary conditions (\ref{eq:mubc}).

\begin{figure}[h]
	\centering
	\includegraphics[scale=.4]{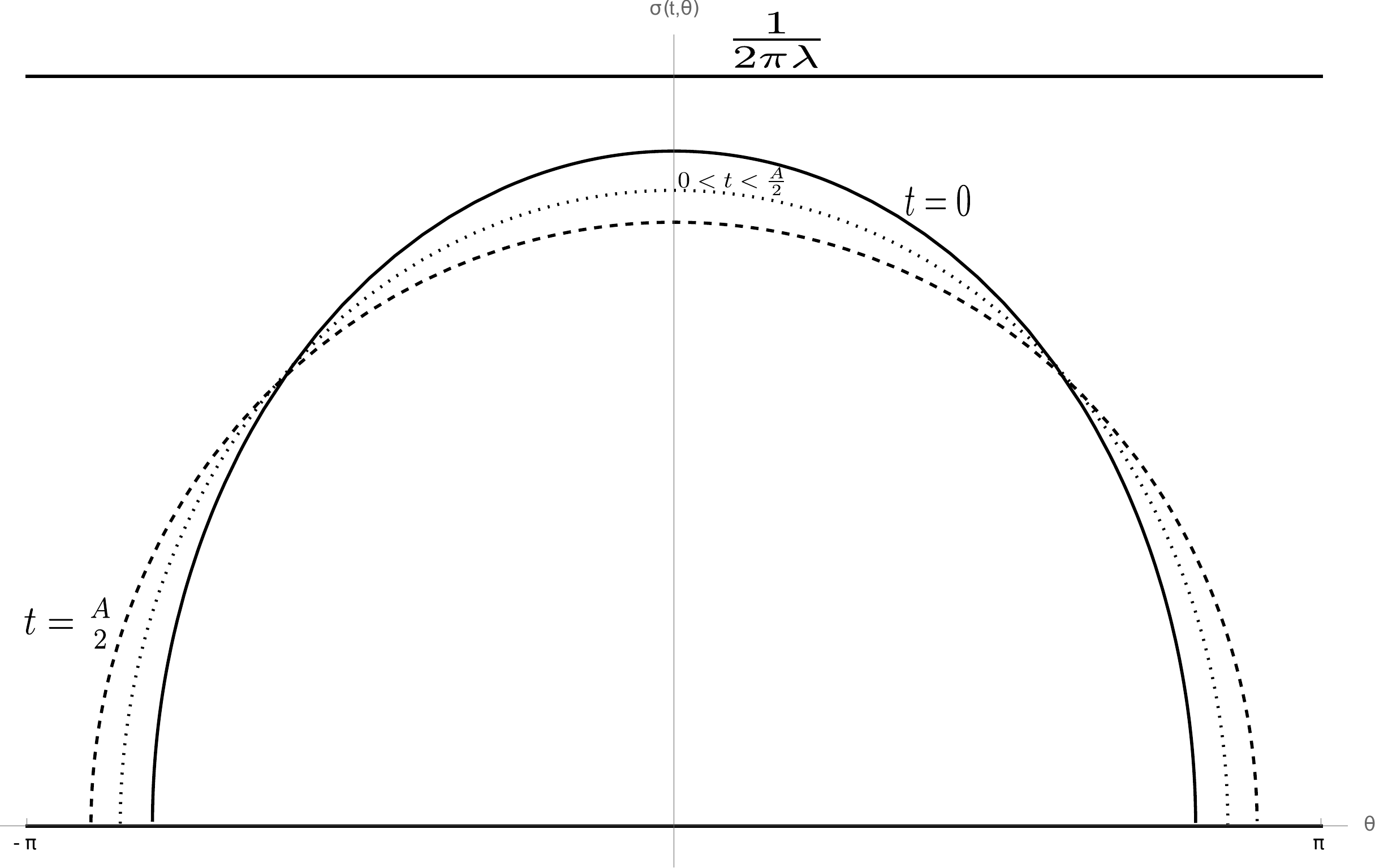}
	\caption{Plot of $\sigma(\theta,t)$ as a function of $\theta$ at $t=0$ (solid), $t=t^*=A/2$ (dashed) and any arbitrary $0<t<t^*$ (dotted).}
	\label{fig:L1equalL2}
\end{figure}

First we consider a special case $L_1=L_2=L$. Solving equation (\ref{eq:mueom}) for symmetric boundary conditions we find
\begin{equation}
    \begin{split}
      \mu(t)= \frac{4L\pi^{2}(8\pi^{2}\lambda^{4}+\sqrt{A^{2}L^{2}\lambda^{4}+64\pi^{4}\lambda^{8}})}{L^{2}(A-t)t+4\pi^{2}(8\pi^{2}\lambda^{4}+\sqrt{A^{2}L^{2}\lambda^{4}+64\pi^{4}\lambda^{8}})}.
    \end{split}
\end{equation}
Since $v(t,\theta)$ is proportional to $\dot \mu(t)$ it is easy to see that the velocity is proportional to $A-2t$ and hence for $L_1=L_2=L$ the fluid equations always admits a solution such that $v(t,\theta)=0$ at $t=A/2$. The density $\sigma(t,\theta)$ coincide with $\sigma_1(\theta)$ at $t=0$ and then starts spreading out as $t$ increases. The spreading is maximum at $t=A/2$. Finally the density again starts contracting and finally becomes $\sigma_2$ at $t=A$. See figure \ref{fig:L1equalL2}. 

If the maximum spread of $\sigma(t,\theta)$ at $t=A/2$ is less than $2\pi$ then $\cW_{\cR_1\cR_2}$ do not observe any phase transition. For
\ben
L\geq \frac{16\pi^{2}\lambda^{2}}{\pi^2-4p^2\lambda^2}
\een
$\sigma(A/2,\theta)$ always has a gap. Since there exists a point $t^*=A/2$ when velocity is zero, there exists a real dominant Young diagram which maximises $\cW_{\cR_1\cR_2}$. It is given by the (\ref{eq:identity1}) and has the form
\begin{equation}\label{eq:domYL}
   \rho(y)=\frac{\sqrt{(8\pi^{2}\lambda^{2}+\sqrt{A^{2}L^{2}+64\pi^{4}\lambda^{4}})(4L-16\pi^{2}\lambda^{2}y^{2})-A^{2}L^{2}y^{2}}}{2L\pi}.
\end{equation}
Since $\sigma^*(\theta)$ has gap, the dominant Young diagram does not saturate the upper bound. However at $L= \frac{16\pi^{2}\lambda^{2}}{\pi^2-4p^2\lambda^2}$ the gap in $\sigma^*(\theta)$ vanishes and the corresponding dominant Young diagram touches the upper cap. Therefore $\cW_{\cR_1\cR_2}$ undergoes a Douglas-Kazakov type phase transition at this critical value of $L$.

The value of the knot invariant can be obtained by calculating the Hamiltonian on this solution and integrating over $A$. After an explicit calculation we obtain the following result (substituting $A=2\pi p \lambda$)
\begin{equation}
    \begin{split}
        \mathrm{W}^{(0)}_{LL}(\lambda) & = -\frac{2\pi  \lambda}{p L} + \frac{1}{2} \sqrt{1+\frac{16\pi^{2}\lambda^{2}}{p^2  L^2}} +\frac{\pi p}{12} - \ln 2
         - \frac12 \sinh^{-1}(\frac{p L}{4\pi\lambda}).
    \end{split}
\end{equation}
This is the HOMFLY-PT polynomial in the double scaling limit for Hopf link where two representations associated with two unknots are same and given by Wigner semicircle distributions. In general HOMFLY-PT is a polynomial of two variables $q=e^{2\pi i/(k+N)}$ and $s=q^N$. However in the double scaling limit the variable $q=1$ and $s=e^{2\pi i \lambda}$, hence $\mathrm{W}^{(0)}_{LL}(\lambda)$ is a function of $\lambda$ only. 

Surprisingly, the final expression for the action $S(\sigma_1,\sigma_2|A)$ with $\sigma_{1/2}$, given by the same semicircle distribution of size $L$, is similar to the limit shape of asymptotic Young diagrams \cite{VerKer77,LogShe,Chattopadhyay:2019pkl} up to an analytic continuation in $p$. In particular
$$\frac{4}{\pi}\lb S(u)+1\rb_{u\ra i u} = \hat\Omega(u), \quad \where \quad u = \frac{p L}{2\pi \lambda}$$ and $\hat\Omega(u)$ is the limit shape. This is nothing but an observation, we think.

One can also choose two different representation by considering the boundary conditions $\mu(0)=L_1$ and $\mu(A)=L_2$. The expression for $\mu(t)$ can also be computed exactly. From these expression we see that the existence of dominant Young diagram depends on the values of $L_1, \ L_2$ and $\lambda$. For a given choice of the set $(L_1, L_2, \lambda)$ if the fluid velocity never reaches zero then the there exists no real dominant Young diagram to maximise (\ref{eq:tildeWL2}). However one can calculate $S(\sigma_1,\sigma_2|A)$ from (\ref{eq:hj}) and hence $\mathrm{W}^0_{L_1L_2}$. In this case $\mathrm{W}^0_{L_1L_2}$ does not observe any phase transition. We see that to get a real dominant Young diagram for $L_1>L_2$ the parameters must satisfy $A\ge \frac{8\sqrt{L_{1}-L_{2}}\pi^{2}\lambda^{2}}{L_{2}\sqrt{L_{1}}}$ and for $L_{2}\ge L_{1}$ the relation is given by $A\ge \frac{8\sqrt{L_{2}-L_{1}}\pi^{2}\lambda^{2}}{L_{1}\sqrt{L_{2}}}$. In this case there exists a critical relation between $L_1, L_2$ and $A$ which determines whether the two point function $\cW_{L_1L_2}$ will undergo a phase transition or not. However such condition does not have any handy expression but can be found numerically. One can also compute the knot polynomial in this case.

\subsection{Calculation of partition function}
\label{sec:unknot}

From (\ref{eq:WLonSM}) we see that we get back the partition function of CS theory on Seifert manifold when $\cR_a=0 \ \forall a$ (no box in the Young diagram). The phase structure of this theory in the aforementioned double scaling limit was discussed in \cite{Chakraborty:2021oyq} by directly solving the saddle point equation. As a consistency check, in this subsection we reproduce the same result from the solution of fluid equations.

When $\cR_a=0$ the corresponding $\sigma_a(\theta)$ is given by
\begin{align}\label{eq:sigmanobox}
	\sigma_a(\theta)= 
	\begin{cases}
		\frac{1}{2\pi \lambda} & \text{for} \quad  -\pi \lambda \leq \theta \leq \pi \lambda \\
		\\
		0  & \text{for} \quad \text{otherwise}.
	\end{cases}
\end{align}
We expect that the solutions of the fluid equations (\ref{eq:eu}) with the boundary conditions that $\sigma(t,\theta)$ merges with (\ref{eq:sigmanobox}) at $t=0$ and $t=A$ will admit an intermediate time $t^*=A/2$ (follows from symmetry) when fluid velocity is zero and the functional inverse of the fluid density at $t^*$ will give the dominant Young diagram representation obtained in \cite{Chakraborty:2021oyq}.

In order to check our expectation we use the dominant Young diagram found in \cite{Chakraborty:2021oyq}
\ben
\rho(y) = \frac{p}{\pi} \tanh^{-1} \lB \sqrt{1-\frac{e^{-\frac{2\pi \lambda}{p}}}{\cosh(\pi \lambda y)}}\rB,
\een
for $0\leq \lambda \leq \frac{p}{\pi}\log[\cosh{(\pi/p)}]$.
Inverse of this function gives $\sigma^*(\theta)$. We now use the identity (\ref{eq:in}) to find $v(t,\theta)$ and $\sigma(t,\theta)$. Since both velocity and density are real functions of $t$ and $\theta$, we solve the real and imaginary parts of this equation and find that $\sigma(t,\theta)$ matches with (\ref{eq:sigmanobox}) at $t=0$ and $t=A$. For $\lambda > \frac{p}{\pi}\log[\cosh{(\pi/p)}]$ the dominant Young diagram distribution has a cap. It is difficult to invert that distribution to find $\sigma^*(\theta)$. However, one can numerically check that other phase also solves the fluid equations.

\section{\texorpdfstring{$Sp(N)$}{SpN} Chern-Simons theory and knot invariants}
\label{sec:spn}

A symplectic group $Sp(n,F)$ is a group of $2n\times 2n$ dimensional symplectic matrices over a field $F$ under matrix multiplication. A $2n\times 2n$ dimensional matrix $A$ is a symplectic matrix if it satisfies the relation
\ben
A^T \Omega_{2n} A = \Omega_{2n}
\een
where $\Omega_{2n}$ is a $2n\times 2n$ dimensional skew-symmetric matrix : $\Omega_{2n}^T+\Omega_{2n}=0$. A standard choice of $\Omega_{2n}$ is 
\ben
\Omega_{2n} = \begin{pmatrix}
0 & \mathbb{I}_n\\
-\mathbb{I}_n & 0
\end{pmatrix}.
\een
Irreducible representations of $Sp(n,F)$ are characterised by Young diagrams with maximum $n$ number of rows. 

CS theory for $Sp(N)$ gauge group is well studied. The large $N$ limit of these theories and their connections with dual string theories were studied in \cite{Sinha:2000ap}. In this section we shall discuss the phase structure of the theory in the aforementioned double scaling limit. Then we show that the analysis, given in section \ref{sec:largeN}, can be extended for $Sp(N)$ gauge group to obtain the invariant for the Hopf link in terms of the collective field theory action. The analysis is little different than that of a $U(N)$ theory since the modular transformation matrices $\cS$ and $\cT$ for $sp(N)_k$ affine algebra have different forms.

\subsection{The partition function}
\label{sec:spPF}

The partition function of CS theory with gauge group $Sp(N)$ and level $k$ on Seifert manifold is given by
 \begin{equation}
     \mathcal{Z}[S^3/\mathbb{Z}_p, Sp(N),k]=\sum_{\cR}\cT_{\cR\cR}^{-p}\cS_{0\cR}^{2}
 \end{equation}
here $\cS$ and $\cT$ for $sp(N)_k$ affine lie algebra are given by
 \begin{equation}
     \begin{split}
         \cS_{\cR\cR'}=(-i)^{\frac{N(N-1)}{2}}\Big(\frac{2}{N+k+1}\Big)^{\frac{N}{2}}\det\left |\sin\Big(\frac{\pi f_{i}(R)f_{j}(R')}{N+k+1}\Big)\right |_{i,j=1}^{N}
     \end{split}
 \end{equation}
where
 \begin{equation}
     f_{i}(\cR)=n_{i}(\cR)-i+N+1 \equiv h_{i}(\cR)
 \end{equation}
and
\begin{equation}
    \mathcal{T}_{\cR \cR'}=\exp\lB {-\frac{i \pi N(N+1)}{12}} + {\frac{i\pi}{2(N+k+1)}\sum_{i=1}^{N}h_{i}(\cR)^{2}}\rB \delta_{\cR,\cR'}.
\end{equation}
We define new variables $\theta_i$
\begin{equation}
    \theta_{i}(\cR)=\frac{\pi h_{i}(\cR')}{N+K+1}=\pi\lambda \frac{h_{i}(\cR)}{N}, \quad \where \ \lambda = \frac{N}{N+k+1}.
\end{equation}
The double scaling limit is given by as before : $N\ra \infty,\ k\ra \infty$ keeping $\lambda$ fixed. For an integrable representation $\cR$ we have $0\leq h_i(\cR)\leq N+k+1$, hence the new variables $\theta_i(\cR)$ satisfies
\ben
0\leq \theta_i(\cR) \leq \pi. 
\een
Although $\theta_i \geq 0$ for a $sp(N)$ representations, we introduce a distribution function $\sigma^{sp}(\theta)$ which defines a symmetric distribution of eigenvalues between $-\pi$ and $\pi$.
\begin{equation}\label{eq:spdensity}
     \sigma^{sp}(\theta)=\frac{1}{2N}\sum_{i=1}^{N}\delta(\theta-\theta_{i}) + \frac{1}{2N} \sum_{i=1}^{N} \delta(\theta+\theta_{i}).
 \end{equation}
Introducing mirror images of the eigenvalues
\be\label{eq:theta-i}
\theta_{-i} = -\theta_i
\ee
$\sigma^{sp}(\theta)$ can be written as
\ben\label{eq:spdensity2}
     \sigma^{sp}(\theta)=\frac{1}{2N}\sum_{{i=-N}\atop{i\neq 0}}^{N}\delta(\theta-\theta_{i}).
\een
Hence $\sigma^{sp}(\theta)$ is a distribution of $2N$ eigenvalues : $N$ $\theta_i$s and their mirror images $\theta_{-i}$s. We should remember the relation (\ref{eq:theta-i}) while taking derivative with respect to $\theta_i$. We also note that $\sigma^{sp}(\theta)$ has upper-cap given by
\ben\label{eq:spgmaspcap}
\sigma^{sp}(\theta) \leq \frac{1}{2 \pi \lambda}
\een
similar to the $U(N)$ case (\ref{eq:capcondition}). 

The partition function in the continuum limit is therefore given by
\begin{equation}
     \begin{split}
      \mathcal{Z}[S^3/\mathbb{Z}_p, Sp(N),k]=\int[d\theta]\exp\lB {-\frac{N^2}{\lambda^2} S_{eff}[\sigma^{sp}]}\rB
     \end{split}
 \end{equation}
 where
 \begin{equation}\label{sadsp}
     \begin{split}
         S_{eff}[\sigma^{sp}] = & - \lambda^{2}\int_{-\pi}^{\pi}\int_{-\pi}^{\pi}\sigma^{sp}(\theta)\sigma^{sp}(\theta') \log\Big[4 \sin^{2}\Big(\frac{\theta- \theta'}{2}\Big) \Big]d\theta d\theta'\\
         & +\frac{2 p\lambda}{\pi}\int_{-\pi}^{\pi} \sigma^{sp}(\theta)\Big(\frac{\theta^{2}}{4}-\frac{\pi^{2}}{12}\Big)+\frac{2p\pi\lambda(1-\lambda)}{12} .
     \end{split}
 \end{equation}
The saddle point equation is given by
\ben
\dashint_{-\pi}^{\pi}\sigma^{sp}(\theta')\cot\Big(\frac{\theta-\theta'}{2}\Big)d\theta'=\frac{p}{2\pi\lambda}\theta .
\een
Thus we see that the saddle point equation for $Sp(N)$ CS theory on $S^3/\mathbb{Z}_p$ is the same as that of $U(N)$ CS theory (\ref{eq:sad}). Hence in the large $N$ limit the phase structure of these two theories are identical. Therefore, in the double scaling limit,  $Sp(N)$ CS theory on $S^3/\mathbb{Z}_p$ admits a third order phase transition at $\lambda = p{\log(\cosh\frac{\pi}{p})}/ {\pi}$.

\subsection{Two point correlator and Hopf link invariants}
\label{sec:spncorrel}

Though the structure of the modular $\cS$ and $\cT$ matrices for $sp(N)_k$ affine algebra is different than that for $u(N)_k$, the two point correlator for the Hopf link admits a description in the language of incompressible fluid and hence can be written in terms of the on shell action of a free collective field theory. The calculation follows the similar line as what we did for $U(N)$ theory but the intermediate steps are different since hook numbers of the Young diagrams of $sp(N)$ representations are always positive. Bulk of the calculations are given in appendix \ref{app:spncalculations}, here we outline the main steps.

Using the expression of modular $\cS$ and $\cT$ matrices we write down the modified two point correlation function (\ref{eq:modifiedWL}) as,
\begin{equation}
\begin{split}
    \widetilde \cW_{\cR_1\cR_2}(S^3/\mathbb{Z}_p,Sp(N),N,k) & = \frac{e^{\frac{i p\pi N(2N+1)}{12}}}{2^{2N(N-1)}}   \sum_{y_{i}(\cR)} \lb \frac{\det[\sin(N\theta_{j}(\cR_{1})y_{i}(\cR))]}{D[\theta(\cR_{1})]}\rb  \\  & \quad \times \lb \frac{\det[\sin(N\theta_{j}(\cR_{2})y_{i}(\cR))]}{D[\theta(\cR_{2})]}\rb 
    e^{-\frac{i\pi p\lambda N}{2}\sum_{i=1}^{N}y_{i}(\cR)^{2}} 
\end{split}
\end{equation}
where
\ben
y_{i}(\cR)=\frac{h_{i}(\cR)}{N}
\een
and
\begin{equation}
    D[\theta(\cR_{1})]=\prod_{i=1}^{N}\sin(\theta_{i}(\cR_{1}))\prod_{i<j}^{N}\sin\Big(\frac{\theta_{i}(\cR_{1})+\theta_{j}(\cR_{1})}{2}\Big)\sin\Big(\frac{\theta_{i}(\cR_{1})-\theta_{j}(\cR_{1})}{2}\Big).
\end{equation}
We follow the same procedure, what we did for $U(N)$ CS theory and define (after a wick rotation $p\to -ip $) $\tilde\cZ_{\cR_1\cR_2}(S^3/\mathbb{Z}_p,Sp(N),N,k)$
\begin{equation}
   \widetilde \cW_{\cR_1\cR_2}(S^3/\mathbb{Z}_p,Sp(2N),N,k)=    \frac{1}{2^{2N(N-1)}} e^{\frac{ p\pi N(2N+1)}{12}} \widetilde \cZ_{\cR_1\cR_2}(S^3/\mathbb{Z}_p,Sp(2N),N,k).
\end{equation}
Following (\ref{eq:FNdef}), in the large $N$ limit we define a similar function function $F(\sigma^{sp}_{1/2},A)$ where $\sigma^{sp}_{1/2}(\theta)$ are eigenvalue distributions corresponding to $\cR_{1/2}$ respectively. Segregating the pure $\sigma^{sp}_{1/2}$ dependent part from $F(\sigma^{sp}_{1/2},A)$, we define a function $S(\sigma^{sp}_{1/2},A)$ given in (\ref{eq:FSrelspN})
and show that $S[\sigma_{1/2}^{sp},A]$ satisfies,
\begin{equation}\label{eq:Ham-JcSpN}
\begin{split}
   \frac{\partial S}{\partial A}=\frac{1}{2} \int{\sigma}^{sp}_{a}\theta)\Big[\Big(\frac{1}{2}\frac{\partial}{\partial\theta}\frac{\delta S}{\delta{\sigma}^{sp}_{a}(\theta)}\Big)^{2}-4\frac{\pi^{2}}{3}{\sigma}^{sp}_{a}(\theta)^{2}\Big]d\theta
\end{split}
\end{equation}
where,
\ben
A= \pi p \lambda .
\een
Thus we see that $S[\sigma^{sp}_a, A]/4$ (Hopf link invariant in $Sp(N)$ CS theory) satisfies the same Hamilton-Jacobi equation and hence the saddle point is governed by an $Sp(N)$ free collective field theory equations.

The real dominant Young diagram in the large $N$ limit can also be obtained by studying the Hamilton's equations. The real dominant representation, if it exists, is given by the inverse of $Sp(N)$ fluid density $\sigma^{sp}(\theta,t)$ at some intermediate time when the fluid velocity is zero. The detailed calculation for dominant representation is given in appendix \ref{app:domrepspn}.

\section{Discussion}
\label{sec:disc}

In this paper we find that the computation of two point correlation function in $U(N)$ and $Sp(N)$ CS theory in $S^3/\mathbb{Z}_p$ (which renders invariant for the Hopf link) in the large $N$ limit boils down to finding solutions of continuity and Navier-Stokes equations of an incompressible one dimensional fluid evolving from $t=0$ to $t=A$ with the initial and final densities corresponding to the representations $\cR_1$ and $\cR_2$. The Hopf link invariant $\mathrm{W}^{(0)}_{\cR_1\cR_2}(\lambda)$ in the large $N$ limit satisfies the Hamilton-Jacobi equation where $p \lambda$ plays the role of time and Hamiltonian is given by $U(N)$ (or $Sp(N)$) free collective field theory Hamiltonian. The invariant for the Hopf link turns out to be equal to the on shell action. Using the method developed in \cite{RamaDevi:1992np} we finally
show that invariants for other torus knots (2,$k$=odd) and links (2,$k$=even) can be obtained from the invariants for the Hopf link and unknot.

We further discuss the large $N$ phase structure of two point correlators in CS theory. Whether two point function undergoes a phase transition or not depends on the evolution of the fluid. The absolute value of the fluid velocity at a given point $\theta$ decreases with time as the fluid starts evolving from $t=0$. The existence of a real dominant representation depends on whether the absolute value of velocity can reach zero at some intermediate time for all $-\pi\leq\theta\leq\pi$. If one start with a class of $\cR_1$ and $\cR_2$ such that corresponding $\sigma_1(\theta)$ and $\sigma_2(\theta)$ are gapped then the density of the fluid starts spreading from its initial distribution $\sigma_1(\theta)$. The spreading is maximum when the velocity is zero (if such solution exists) and then starts shrinking and goes to its final distribution $\sigma_2(\theta)$ at $t = A$. The maximum spreading of fluid density (at $t=t^*$) depends on the choice of $\cR_{1/2}$ and $\lambda$. If the maximum spread lies between $-\pi$ and $\pi$ then there is no phase transition in $\mathrm{W}_{\cR_1\cR_2}^{(0)}$. However, $\mathrm{W}_{\cR_1\cR_2}^{(0)}$ observes a third order phase transition otherwise. The CS theory enjoys \emph{level-rank} duality. A theory with rank $N$ and level $k$ is dual to a theory with level and rank exchanged. The two point correlation function in the dual theory also admits the similar fluid structure. It would be interesting to understand the relation between the fluid and its dual fluid in the large $N$ limit.

The study of large $N$ correlation functions is important in many aspects. It sets up the platform to check the generalised volume conjecture \cite{Kashaev,Murakami,Hitoshi1,Gukov:2003na}. Further, based on the Gopakumar-Vafa conjecture \cite{Gopakumar:1998ki} it was observed in \cite{Ooguri:1999bv} that the CS invariants are mapped to topological string amplitudes on Riemann surfaces with boundaries in the topological string theory side. The knot invariants were reformulated in terms of new invariants (integer invariants) capturing the spectrum of M2 branes ending on M5 branes. The results were checked explicitly for unknot. Large $N$ analysis of this observation and its generalisation to other links were considered in \cite{Labastida:2000yw}. Our analysis to compute invariants for a class of knots and links will be useful to further investigate the connection between CS theory and topological string theory beyond partition function. 

As we mentioned in section \ref{sec:UNCScorrelators} that in $S^3$ there exists a canonical framing $\cK = \cS$. In this framing the two point correlation function for the Hopf link is given by $\cS_{\cR_1\cR_2}$ and does not show phase transition for any $\cR_{1/2}$. In Seifert framing the same quantity is given by (\ref{eq:W2}). The sum is over the integrable representations. Using the properties of $\cS$ and $\cT$ matrices one can show that these two expressions are related to each other up to a phase factor. The question is why we see a phase transition in two point function in the double scaling limit. If one takes the $N, k \ra \infty $ limit without any restriction, then the sum in (\ref{eq:W2}) runs over all possible Young diagrams with any number of rows and any number of columns. However, here we are considering a particular limit $N, k \ra \infty $ keeping $N/k$ fixed. Under this condition the sum becomes restricted - one does not sum over all possible Young diagrams. Therefore we do not expect that in the double scaling limit the above equality holds. This was also the reason behind the phase transition in CS theory in $S^3$ studied in \cite{Chakraborty:2021oyq}. The framing dependence of correlation functions in CS theory bears a mining in the topological string theory side. Framing is related to inherent ambiguity in the open topological string amplitude related to the IR geometry of the D-brane \cite{Aganagic:2001nx,Marino:2001re}. It would be interesting to study the topological string amplitudes in the same double scaling limit and understand the connection better. We keep the problem for future.

\vspace{.2cm}


\noindent
{\bf Acknowledgments\ :} The work of SD is supported by the \emph{MATRICS} (grant no.  \emph{MTR/ 2019/ 000390}, the Department of Science and Technology, Government of India). SD acknowledges the Simons Associateship of the Abdus Salam ICTP, Trieste, Italy. We are indebted to people of India for their unconditional support toward the researches in basic science.

\appendix

\section{Framing dependence of partition function}\label{app:surgery}

Partition function of CS theory in three dimensions depends on choice of framing \cite{Wittenjones}. In this appendix we discuss the framing dependence in details. 

Canonical quantisation of CS theory on a three manifold $\cM$ with a boundary $\Sigma$ produces the "physical Hilbert space" $\cH(\Sigma)$. In \cite{Wittenjones} Witten constructed $\cH(\Sigma)$ in terms of conformal blocks of WZW model on $\Sigma$ with a gauge group $G$ and and level $k$. One can explicitly construct the Hilbert space for $\Sigma=S^2$ and $\Sigma= \mathbb{T}^2$. For $\Sigma=S^2$ the Hilbert space is trivial (dimension one). However for $\Sigma=\mathbb{T}^2$ the space of conformal blocks has one-to-one correspondence with integrable representations of $\mathbf{g}_k$. There is a natural choice of basis for $\cH(\Sigma= \mathbb{T}^2)$. The basis vectors are given by integrable representations $\ket{\cR_a}$ of $\mathbf{g}_k$ : these are the states associated with the partition function of CS theory in a solid torus $\mathbf{T}^2$ with a WL in representation $\cR_a$ along the non-contractible cycle. Therefore the Hilbert space is finite dimensional and spanned by these integrable representations.

In order to write the partition function of CS theory on a generic three manifold $\cM$ we split the the manifold into two parts $\cX_L$ and $\cX_R$ sharing a common boundary $\Sigma$. The path integral of CS theory on $\cX_{R}$ corresponds to a vector $\ket{\phi}$ in $\cH(\Sigma)$. Since the boundary $\Sigma$ of $\cX_L$ has an opposite orientation of that of $\cX_R$ the path integral on $\cX_L$ is mapped to a vector $\bra{\psi} \in \cH^*(\Sigma)$ where $\cH^*(\Sigma)$ is dual of $\cH(\Sigma)$. Since the manifold $\cM$ can be obtained by gluing $\cX_L$ and $\cX_R$ along $\Sigma$, the partition function of CS theory on $\cM$ is therefore given by \cite{Wittenjones}
\ben\label{eq:pfonM}
\cZ(\cM) = \langle\psi|\phi\rangle.
\een
Using this result Witten showed that one can write the partition function and correlation functions of CS theory in $S^3$ and other generic three manifolds from the partition function and WLs in $S^2\times S^1$.

In order to understand the prescription in detail let us start with CS theory on a three manifold $\cM$. We consider a WL $\cW_{\cR_a}= \Tr_{\cR_a}U_{\mathrm{K}}$ in $\cM$ along a knot $\mathrm{K}$, where $U_{\mathrm{K}}= P \exp\lb \int_{\mathrm{K}} A\rb$ and $\cR_a$ is an integrable representation of $\mathbf{g}_k$. We take a tubular neighbourhood of the knot $\mathrm{K}$ which is a solid torus $\mathbf{T}^2$ such that $\partial \mathbf{T^2}=\mathbb{T}^2$ is a torus. We take the $\mathbf{T}^2$ out of the manifold and hence the three manifold $\cM$ is now a connected sum of $\cX_R = \mathbf{T}^2$ with a WL inserted and the reminder $\cX_L$. Note both $\cX_L$ and $\cX_R$ have common boundary $\mathbb{T}^2$. Following the work of Verlinde \cite{Verlinde:1988sn} Witten showed that the path integral over $\mathbf{T}^2$ with a WL $\cW_{\cR_a}$ along the non-contractible cycle of $\mathbf{T}_2$ is mapped to a state $\ket{\cR_a}$ in $\cH(\mathbb{T}^2)$. Thus, following (\ref{eq:pfonM}) we see that the expectation value of WL $\cW_{\cR_a}$ in representation $R_a$ in $\cM$ can be written as
\ben
\langle \cW_{\cR_a} \rangle_{\cM} = \braket{\psi}{\cR_a}
\een
where $\bra{\psi} \in \cH^*(T^2)$ is path integral over $\cX_L$. For $\cR_a$ to be a trivial representation, the WL is equal to $1$. Therefor $\braket{\psi}{0}$ is the partition function on $\cM$.

Now, before gluing the solid torus with $\cX_L$ one can also make a diffeomorphism on the boundary of $\mathbf{T}^2$. Such operation (scooping out $\mathbf{T}^2$ from $\cM$ $\ra$ apply diffeomorphism on the boundary $\ra$ gluing back with $\cX_L$) generates a new manifold $\tilde \cM$. Let us first understand this with the help of a simple example. Suppose our $\cM = S^2 \times S^1$. The manifold $S^2\times S^1$ can be written as a connected sum of two solid tori. To understand this one can think that a solid torus is a \emph{disc} times a \emph{circle} : $\mathbf{T}^2=D\times S^1$. When we glue two disc at the boundary we get the two manifold $S^2$. Therefore, when we glue two solid tori along their boundaries without any diffeomorphism we get $S^2\times S^1$. Thus we see that when we scoop out a solid torus from $S^2\times S^1$ the reminder is also a solid torus. Consider now a solid torus $\mathbf{T}^2$ embedded in $S^3$ which is $R^3 \cup \infty$. $\mathbf{T}^2$ is invariant under inversion. The exterior of this $\mathbf{T}^2$ is another solid torus $\mathbf{T'}^{2}$ (as we have identified the points at infinity). However, there is a difference. The contractible cycle in $\mathbf{T}^2$ is mapped to the non-contractible cycle in $\mathbf{T'}^2$ and vice versa. Therefore these two tori are related to each other by $S$ modular transformation on the boundary $\mathbb{T}^2$. Thus when we glue $\mathbf{T}^2$ and $\mathbf{T'}^2$ we get $S^3$. Now starting from $\cM = S^2\times S^1$ we split the manifold in two solid tori. Then we perform an $S$ modular transformation (i.e. inverting the torus) on the boundary of one of them (say the right one) and then glue them again. This surgery produces $S^3$ from $S^2\times S^1$.

Consider now a CS theory in a solid torus with a WL in representation $\cR_a$ along the non-contractible circle. The path integral maps to $\ket{\cR_a} \in \cH(\mathbb{T}^2)$. 
If $K$ is the diffeomorphism that acts on the boundary of the solid torus, the path integral on $\mathbf{T}^2$ changes and hence is mapped to a different state in $\cH(\mathbb{T}^2)$. The new state can be written as
\ben
\ket{\chi}=\sum_{\cR_b} \cK_a^{\ b} \ket{\cR_b}.
\een
The matrix $\cK \in \text{Hom}(\cH(\mathbb{T}^2), \cH(\mathbb{T}^2))$ depends on the diffeomorphism $K$ that acts on the torus. If we now glue this solid torus with $\cX_L$ we get a new manifold $\tilde\cM$ with a WL in $\cR_a$. The expectation value of the WL in representation $\cR_a$  in $\tilde \cM$ can be written in terms of expectation value of WLs in $\cM$
\ben
\langle W_{R_a}\rangle_{\tilde \cM} = \sum_{\cR_b} \cK_a^{\ b} \langle W_{\cR_b} \rangle_{\cM}.
\een
Hence the partition function on $\tilde \cM$ can be obtained by considering $R_a$ to be a trivial representation
\ben\label{eq:pfmtilde}
Z(\tilde \cM) = \sum_{\cR_b} \cK_0^{\ b} \langle W_{R_b} \rangle_{\cM}.
\een
This is a very powerful relation.

The partition function of \cs theory on $S^2\times S^1$ can be calculated using (\ref{eq:pfmtilde}). As explained above the manifold $S^2\times S^1$ can be written as a connected sum of two solid tori. As mentioned earlier, partition function in $\mathbf{T}^2$ without any WL is mapped to $\ket{0}$ in $\cH(\mathbb{T}^2)$. Thus we get partition function of CS theory on $S^2\times S^1$ is given by
\ben\label{eq:pfs2s1}
Z(S^2\times S^1) = \braket{0}{0}=1.
\een
Similarly when we have a non-trivial WL in $S^2\times S^1$ we can split the manifold into two solid tori with one torus containing the WL along the con-contractible cycle. Hence the expectation value of a WL in $S^2\times S^1$ is given by
\ben
\langle W_{\cR_a} \rangle_{S^2\times S^1} = \braket{0}{\cR_a}=\delta_{0\cR_a}
\een

Our goal is to generate $S^3$ from $S^2\times S^1$ by surgery. Starting from $S^2\times S^1$ we split the manifold in two solid torus. Then we perform an $S$ modular transformation on the boundary $\mathbb{T}^2$ and then glue them again. This surgery produces $S^3$. The diffeomorphism $K$ is the modular transformation $S$ and hence the corresponding $\cK$ matrix is the modular transform matrix $\cS$ in $\cH(\mathbb{T}^2)$. Therefore following (\ref{eq:pfmtilde}) and (\ref{eq:pfs2s1}) we find the CS partition function on $S^3$ is given by
\ben
Z(S^3) = \sum_{\cR_b} \cS_{0\cR_b} \delta_{0\cR_b} = \cS_{00}.
\een
However, instead of choosing $\cK = \cS$ if we choose $\cK=\cT \cS \cT = \cS \cT^{-1} \cS$, this also produces $S^3$ but in a different framing called \emph{Seifert framing}. In this framing the partition function is given by
\ben
Z_{SF}(S^3) = \sum_{\cR} \cS_{0\cR}^2 \cT_{\cR\cR}^{-1}
\een
which is same as $Z(S^3)$ up to a phase.

Partition function for CS theory on a generic Lens space $S^3/\mathbb{Z}_p$ (a Seifert manifold with $g=0$) can be obtained from $S^2\times S^1$ by choosing $\cK = \lb \cT \cS \cT \rb^p = \lb \cS \cT^{-1} \cS \rb^p$. The partition function is given by (\ref{eq:cspfSF}) with $g=0$.

In order to get the partition function of Hopf link in $S^3/\mathbb{Z}_p$ we can start with CS theory on $S^2\times S^1$ with two WLs. We consider a solid torus along one WL and  split the manifold with two parts scooping out the solid torus. The reminder is also a solid torus with the other loop. Partition functions on these two tori are given by $\ket{\cR_a}$ and $\bra{\cR_b}$. Hence CS partition function with two WLs is given by
\ben
Z(S^2\times S^1, \cR_a\cR_b) = \braket{\cR_b}{\cR_a} = \delta_{\cR_a\cR_b}.
\een
Before gluing if we give an inversion on the right torus by choosing $\cK = \lb \cS \cT^{-1} \cS \rb^p$ we get Hopf link in $S^3/\mathbb{Z}_p$ given by (\ref{eq:W2}).

We now want to find the correlation of three WLs in representations $\cR1$, $\cR_2$ and $\cR_3$ in $S^2\times S^1$. We can split $S^2\times S^1$ with three WLs into two solid tori : one contains a WL in $\cR_1$ and the other torus contains two WLs in $\cR_2$ and $\cR_3$. Suppose the CS partition function in $\mathbf{T}^2$ with two WLs along non-contractible circle is mapped to a state $\ket{\cR_2\cR_3}$ in $\cH(\mathbb{T}^2)$. Since $\cR_i$s correspond to primary fields in WZW we have
\ben\label{eq:verlinde}
\ket{\cR_2\cR_3}=\sum_{\cR} \cN^{\cR}_{\cR_1\cR_2}\ket{\cR}
\een
where $\cN^{\cR}_{\cR_1\cR_2}$ is the Verlinde numbers. Hence 
\ben
Z(S^2\times S^1, \cR_1,\cR_2,\cR_3) = \cN^{\cR_1}_{\cR_2\cR_3}.
\een
One can now use the result (\ref{eq:verlinde}) to write the $n$-point correlation function in $S^2\times S^1$ given by (\ref{eq:s2s1correl}) for $g=0$ with the help of Verlinde formula
\ben
\cN^{\cR_1}_{\cR_2\cR_3} = \sum_{\cR}\frac{\cS_{\cR\cR_2}\cS_{\cR\cR_3}\cS^*_{\cR\cR_1}}{\cS_{0\cR}}.
\een
Starting from (\ref{eq:s2s1correl}) we take a solid torus out from $S^2\times S^1$ with any of the $n$ WLs inside, apply an inversion on the torus and then put that back inside. In this process we generate an $S^3$ with the link, shown in fig.\ref{fig:nWL} inside and the invariant is given by (\ref{eq:ncorrelWZW}).

\section{Review of \texorpdfstring{$U(N)_k$}{SD} Chern-Simons theory on Seifert Manifold }
\label{app:pfreview}

The partition function for $U(N)$ CS theory on Seifert manifold is given by
\ben\label{eq:cspfSF}
\cZ_{N,k} = \sum_{\cR} \cS_{0\cR}^{2-2g}\cT_{\cR\cR}^{-p}.
\een
Since the affine Lie algebra $u(N)_k$ is the quotient of $su(N)_k \times u(1)_{N(k+N)}$ by $\mathbb{Z}_N$, the $u(N)$ representations can be expressed in terms of $su(N)$ representations (denoted by $R$) and eigenvalues of $u(1)$ generator $Q$ : $\cR = (R,Q)$, where $Q= r(R)\ \text{mod} \ N$ and $r(R)$ is the number of boxes in $R$. The trivial representation $\cR=0$ means both $R=0$ and $Q=0$. The modular transform matrix $\cS_{\cR\cR'}$ for $u(N)_k$ in terms of $su(N)$ representations and the $u(1)$ charges is given by \cite{Naculich:2007nc, Naculich:1991, yellowbook}
\begin{eqnarray}\label{eq:Smod}
\cS_{\cR\cR'}=\frac{(-i)^{\frac{N(N-1)}2}}{(k+N)^{\frac N2}} e^{-\frac{2\pi i Q Q'}{ N(N+k)}}\det M(R,R')
\end{eqnarray}
where, $M(R,R')$ is a $N\times N$ matrix with elements,
\ben
M_{ij}(R,R') = \exp\lB\frac{2\pi i}{ k+N}\phi_i(R)\phi_j(R')\rB,
\een
\ben\label{eq:phii}
	\phi_i(R) = l_i-\frac{r(R)}{ N}-i+\frac{1}{2} (N+1)
\end{eqnarray}
and $l_i$'s are the number of boxes in $i^{th}$ row in $R$. The modular transformation  matrix $\cT_{\cR \cR'}$ is given by
\ben\label{eq:Trr}
\cT_{\cR \cR'} & =e^{2\pi i(h_{R}-\frac{c}{24})} \delta_{\cR\cR'},\
h_{R} =\frac{1}{2}\frac{C_{2}(\cR)}{k+N},\
c =\frac{N(Nk+1)}{k+N}\ \ \
\een
where $\cC_2(\cR)$ is the quadratic Casimir of $u(N)_k$. The $u(N)$ representations $\cR$ can be characterised by extended YDs by introducing the number of boxes in $i^{th}$ row $n_i=l_i+s$, for $1\leq i\leq N-1$ and $n_N=s$ where $s\in \mathbb{Z}$. $n_i$s can be negative and the corresponding YDs will have \emph{anti-boxes} \cite{Aganagic:2005dh}. In terms of these $n_i$s the quadratic Casimir $C_2(\cR)$ is given by
\ben
C_2(\cR) = \sum_{i=1}^N n_i(n_i-2i+N+1).
\een
\emph{For an integrable representation $\cR$ of $u(N)_k$}
\ben
\label{eq:intrepdef}
-\frac{k}{2}\leq n_N\leq \cdots \leq n_1 \leq \frac{k}{2}.
\een
and hence
\ben
-\frac{k}{2}< h_N < \cdots < h_1 \leq \frac{k}{2} + N-1.
\een

In terms of the variables $\theta_i$ introduced in (\ref{eq:thetaidef}),  the CS partition function (\ref{eq:cspfSF}) in $S^3/\mathbb{Z}_p$ in the continuum limit is given by
\ben\label{eq:se}
\begin{split}
\cZ_{N,k}^{p}   & = \int [d\theta]e^{-(N+K)^{2}S_{eff}[\sigma]}, \quad \where  \\
        S_{eff}[\sigma] & = \frac{p \lambda}{\pi}\int \sigma(\theta)\lb \frac{\theta^{2}}{4} - \frac{\pi^{2}}{12} \rb d \theta + \frac{\pi p \lambda(1-\lambda)}{12} \\ 
        & - \frac{\lambda^{2}}{2}\int \dashint\sigma(\theta)\sigma(\theta')\log\lB 4\sin^{2}\lb \frac{\theta-\theta'}{2}\rb \rB d\theta d\theta'.
        \end{split}
\een
The saddle point equation for $\sigma(\theta)$, obtained from the effective action is given by
\begin{equation}\label{eq:sad}
          \dashint\sigma(\theta')\cot\left(\frac{\theta-\theta'}{2}\right)d\theta'=\frac{p}{2\pi\lambda}\theta .
\end{equation}
In the large $N$ limit we have to solve this equation for $\sigma(\theta)$ in presence of this constraint (\ref{eq:capcondition}).

The unitary matrix model (\ref{eq:se}) has a gapped phase in the large $k,N$ limit and the eigenvalue distribution is given by \cite{Chattopadhyay,Chattopadhyay:2019lpr},
\begin{eqnarray}\label{eq:csev1gap}
\sigma(\theta) =
\frac{p}{2\pi^2 \l}\tanh^{-1}\left[ 
\sqrt{1- \frac{e^{-\frac{2\pi\l}{p}}}{\cos^{2}\frac{\theta}{2}}}\right], \quad \text{for} \ \ -2\cos^{-1}e^{-\frac{\pi \l}{p}}<\theta< 2\cos^{-1}e^{-\frac{\pi\l}{p}}.
\end{eqnarray}
It turns out that the solution saturates the upper bound (\ref{eq:capcondition})  at $\lambda= p/\pi \log\cosh(\pi/p) \equiv \lambda^* $ \cite{Chattopadhyay:2019lpr} and is not valid beyond that.

\emph{Cap-gap phase :} It was observed in \cite{Chakraborty:2021oyq} that for $\lambda>\lambda^*$ the saddle point equation admits a \emph{cap-gap} solution. The eigenvalue density $\sigma(\theta)$ for $\lambda > p/\pi \log\cosh(\pi/p)$ is given by
 \ben
   \sigma(\theta) = \Bigg\{ {\frac{1}{2\pi \lambda}  \hspace{.3cm} \text{for}\  -\theta_{2}<\theta<\theta_{2} \hfill \atop
     \hat\sigma(\theta)\ \ \text{for}\  -\theta_{1}<\theta<-\theta_{2}\  \text{and}\  \theta_{2}<\theta<\theta_{1}. }
 \een
where,
\ben
\hat{\sigma}(\theta) &=&  \frac{|\sin\phi|}{\pi^{2}\lambda}\frac{\sqrt{(\sin^{2}\frac{\phi}{2}-\sin^{2}\frac{\theta_{2}}{2})(\sin^{2}\frac{\theta_{1}}{2}-\sin^{2}\frac{\phi}{2})}}{\sqrt{(1+\cos\theta_{2})(1-\cos\theta_{1})}}\Bigg[ \frac{4\lb \Pi(n_{2},m_{2})-\sin^{2}\frac{\phi}{2}K(m_{2})\rb}{\sin^{2}\phi} \nonumber \\ 
&& \qquad \frac{2p \lb \cos^{2}\frac{\theta_{1}}{2} \Pi(\psi,n_{1},m_{1}) - \cos^{2}\frac{\phi}{2} F(\psi,m_{1})\rb} {(1+\cos\phi) (\cos\phi-\cos\theta_{1})} \Bigg].
\een
The constants $m_1,m_2,n_1,n_2, \psi$ are given in \cite{Chakraborty:2021oyq}.

\section{Explicit derivation of Hamilton-Jacobi equation}\label{app:Fcalculation}

\subsection{\texorpdfstring{$U(N)$}{UN} Chern-Simons theory}\label{app:uncalculations}

Replacing $\widetilde{\cZ}_{\cR_{1}\cR_{2}}=e^{N^{2}F_{N}}$ in (\ref{eq:den}) and using the relation \begin{equation}
\begin{split}
\frac{1}{D[\theta^{(a)}]}\frac{\partial^{2}D[\theta^{(a)}]}{\partial\theta_{k}^{(a)^{2}}}=N\frac{\partial U_{k}}{\partial\theta_{k}^{(a)}}+N^{2}U_{k}^{2}
\end{split}
\end{equation}
for
\begin{equation}
\begin{split}
U_{k}=\frac{1}{N}\frac{\partial\log[D[\theta^{(a)}]}{\partial\theta_{k}^{(a)}}=\frac{1}{2N}\sum\limits_{\substack{j=1 \\ j\neq k}}^N \cot\Big(\frac{\theta_{k}^{(a)}-\theta_{j}^{(a)}}{2}\Big)
\end{split}
\end{equation}
we get
\begin{equation}\label{eq:eff}
\begin{split}
2\frac{\partial F_{N}}{\partial A}=\frac{1}{N}\sum_{k=1}^{N}\frac{\partial^{2} F_{N}}{\partial\theta_{k}^{(a)^{2}}}+\frac{1}{N}\sum_{K=1}^{N}\Big(N\frac{\partial F_{N}}{\partial\theta_{k}^{(a)}}\Big)^{2}
+\frac{2}{N}\sum_{k=1}^{N}U_{k}N\frac{\partial F_{N}}{\partial\theta_{k}^{(a)}}+\frac{1}{N^{2}}\sum_{k=1}^{N}\frac{\partial U_{k}}{\partial\theta_{k}^{(a)}}+\frac{1}{N}\sum_{k=1}^{N} U_{k}^{2}
\end{split}
\end{equation}
Calculating
\begin{equation}
\begin{split}
\frac{1}{N^{2}}\sum_{k=1}^{N}\frac{\partial U_{k}}{\partial\theta_{k}^{(a)}}=-\frac{1}{4N^{3}}\sum_{k=1}^{N}\sum\limits_{\substack{j=1 \\ j\neq k}}^N\frac{1}{\sin^{2}\Big(\frac{\theta_{k}^{(a)}-\theta_{j}^{(a)}}{2}\Big)}
\end{split}
\end{equation}
we note that the right hand side is \emph{zero} in the large $N$ limit when $\theta_{k}^{(a)}\neq\theta_{j}^{(a)}$. It only gives a nonzero contribution for $\theta_{k}^{(a)}\approx\theta_{j}^{(a)}$. In the large $N$ limit we define continuous distribution functions $\sigma_a(\theta)$ for $\{\theta_i^{(a)}\}$ given by (\ref{eq:sigmadef})
\ben
\theta_i^{(a)} = \theta^{(a)}(x).
\een
 Thus $\theta_{k}^{(a)}-\theta_{j}^{(a)}\approx\frac{|k-j|}{N\sigma_{(a)}(\theta_{k})}$, also in this limit the \emph{sum} will be replaced by integration and all the partial derivatives by corresponding functional derivatives
\begin{equation}
\begin{split}
\frac{1}{N}\sum_{k=1}^{N}\to\int\sigma_{a}(\theta)d\theta;\\
N\frac{\partial}{\partial\theta_{k}^{(a)}}\to\frac{\partial}{\partial\theta}\frac{\delta}{\delta\sigma_{a}(\theta)}
\end{split}
\end{equation}
Finally using the identity
$$\sum\limits_{\substack{j=1 \\ j\neq k}}^{N\to\infty}\frac{1}{(j-k)^{2}}=\frac{\pi^{2}}{3}$$
the equation (\ref{eq:eff}) reduced to (\ref{eq:free1}). Here we have neglected the term $\frac{\partial^{2} F_{N}}{\partial\theta_{k}^{(a)^{2}}}$ as it is $\mathcal{O}(\frac{1}{N})$ in the large $N$ limit. 

\subsection{\texorpdfstring{$Sp(N)$}{SpN} Chern-Simons theory}\label{app:spncalculations}

The derivation of the Hamilton-Jacobi equation for $Sp(N)$ CS theory falls in the same line as that of a $U(N)$ theory once we define the eigenvalue density for the eigenvalues and their mirror images (\ref{eq:spdensity2}).

The function $\widetilde \cZ_{\cR_1\cR_2}(S^3/\mathbb{Z}_p,Sp(N),N,k)$ is given by,
\be
\label{eq:zforsp}
\begin{split}
    \widetilde \cZ_{\cR_1\cR_2}(S^3/\mathbb{Z}_p,Sp(N),N,k) & = \sum_{y_{i}(\cR)}\frac{\det[\sin(N\theta_{j}(\cR_{1})y_{i}(\cR))]}{D[\theta(\cR_{1})]}\\
    & \frac{\det[\sin(N\theta_{j}(\cR_{2})y_{i}(\cR))]}{D[\theta(\cR_{2})]}e^{-\frac{A N}{2}\sum_{i=1}^{N}y_{i}(\cR)^{2}} .
    \end{split}
\ee
the quantity $\widetilde \cZ_{\cR_1\cR_2}$ in (\ref{eq:zforsp}) will satisfy (\ref{eq:den}) with $A=\pi p\lambda$. The quantity $\log[D[\theta]]$ for $Sp(N)$ theory has the form
\begin{equation}
 \log[D[\theta]]=\frac{1}{4}\sum_{i=-N}^{N}\log[\sin\theta_{i}]+\frac{1}{4}  \sum_{k=-N}^{N}\sum\limits_{\substack{j=-N\\j\neq k}}^{N}\log\Big[\sin\Big(\frac{\theta_{k}-\theta_{j}}{2}\Big)\Big]  . 
\end{equation}
For $Sp(N)$ theory we define the quantity $U_k$ as follows,
\begin{equation}
    U_{k}=\frac{1}{N}\frac{\partial \log[D[\theta^{(a)}]]}{\partial\theta_{k}^{(a)}}=\frac{1}{2N}\cot\theta_{k}^{(a)}+\frac{1}{2N}\sum\limits_{\substack{j=-N\\j\neq k}}^{N}\cot\Big(\frac{\theta_{k}^{(a)}-\theta_{j}^{(a)}}{2}\Big)
\end{equation}
and find that it satisfies
\begin{equation}
    \frac{1}{N^{2}}\frac{\partial U_{k}}{\partial\theta_{k}^{(a)}}=-\frac{1}{4N^{3}}\sum\limits_{\substack{j=-N\\j\neq k}}^{N}\frac{1}{\sin^{2}(\frac{\theta_{k}^{(a)}-\theta_{j}^{(a)}}{2})}-\frac{3}{4N^{3}}\frac{1}{\sin^{2}\theta_{k}}
\end{equation}
In the large $N$ limit the only contribution will comes from those $\theta_{j}$ which are close to $\theta_{k}$. Thus $\theta_{k}^{(a)}-\theta_{j}^{(a)}\approx\frac{|k-j|}{2N\hat{\sigma}_{(a)}(\theta_{k})}$ and using the identity $\lim_{N\to\infty}\sum\limits_{\substack{j=-N\\j\neq k}}^{N}\frac{1}{(j-k)^{2}}=\frac{\pi^{2}}{3}$ we find that
\begin{equation}\label{u for sp}
    \begin{split}
        \frac{1}{N}\sum_{k=1}^{N}\frac{\partial U_{k}}{\partial\theta^{(a)}_{k}}\approx-\frac{1}{N^{2}}\sum_{k=1}^{N}4N^{2}\frac{\pi^{2}}{3}\hat{\sigma}_{(a)}(\theta_{k}).
    \end{split}
\end{equation}
Assuming $\widetilde \cZ_{\cR_1\cR_2}$ is dominated by a single representation in the large $N$ limit we use the ansatz $\widetilde \cZ_{\cR_1\cR_2}=e^{N^{2}F_{N}}$,where $F_{N}$ can be written
\begin{equation}
    F_{N}=S_{N}-\frac{1}{N^{2}}\sum_{a=1}^{2}\Big[\frac{1}{4}\sum_{i=-N}^{N}\log[\sin\theta_{i}^{(a)}]+\frac{1}{4}  \sum_{k=-N}^{N}\sum\limits_{\substack{j=-N\\j\neq k}}^{N}\log\Big[\sin\Big(\frac{\theta_{k}^{(a)}-\theta_{j}^{(a)}}{2}\Big)\Big]  \Big]
\end{equation}
from above equation we can find two relation
\begin{eqnarray}\label{eq:FSrelspN}
\begin{split}
    \frac{\partial F_{N}}{\partial\theta_{k}^{(a)}}=\frac{\partial S_{N}}{\partial\theta_{k}^{(a)}}-\frac{1}{N}U_{k}\\
    \frac{\partial F_{N}}{\partial A}=\frac{\partial S_{N}}{\partial A}
\end{split}
\end{eqnarray}
Using the above two relation we can reduce the differential equation of $\widetilde \cZ_{\cR_1\cR_2}$ to differential equation of $S_{N}$
\begin{equation}\label{s for sp}
    \frac{\partial S_{N}}{\partial A}=\frac{1}{2N}\sum_{K=1}^{N}\Big[\Big(N\frac{\partial S_{N}}{\partial\theta_{k}^{a}}\Big)^{2}+\frac{1}{N}\frac{\partial U_{k}}{\partial\theta_{k}}\Big]
\end{equation}
In the continuum we assume that the density converges to a smooth function and the sum and partial derivative replace by
\begin{equation}
    \lim_{N\to\infty}\frac{1}{2N}\sum_{j=-N}^{N}\to\int\sigma^{sp}_{(a)}(\theta)d\theta \qquad 2N\frac{\partial}{\partial\theta_{k}^{(a)}}\to\frac{\partial}{\partial\theta}\frac{\delta}{\delta{\sigma}^{sp}_{(a)}(\theta)}|_{\theta=\theta_{k}^{(a)}}
\end{equation}
Replacing the sum by integral and using (\ref{u for sp}) we get the continuum version of (\ref{s for sp}) as given by (\ref{eq:Ham-JcSpN}). With redefinition of $S=4S'$ the above equation is same as (\ref{eq:hj}) which can be mapped to a Hamilton-Jacobi equation with Hamiltonian
\begin{equation}
    H[\sigma,\Pi]=\frac{1}{2}\int\sigma(\theta)^{2}\Big[\Big(\frac{\partial\Pi}{\partial\theta}\Big)^{2}-\frac{\pi^{2}}{3}\sigma(\theta)^{2}\Big]d\theta
\end{equation}
the Hamilton's equation of motion is same as (\ref{eq:eu}) with the boundary condition $\sigma(t=0,\theta)=\hat{\sigma}_{1}(\theta)$ and $\sigma(t=A,\theta)=\hat{\sigma}_{2}(\theta)$

\section{From eigenvalue density to Young tableau density}\label{app:evtoYD}

To find the the dominant Young tableau density we can use the large $N$ properties of the Itzykson-Zuber integral

\begin{equation}
    I_{N}(A,B)\equiv\int[dU]e^{N\Tr(AUBU^{\dagger})}=\frac{\det||e^{N a_{k}b_{j}}||}{\Delta(a)\Delta(b)}.
\end{equation}
Here $A$ and $B$ are arbitrary hermitian matrices, $a_{k}$ and $b_{j}$ are their eigenvalues and $\Delta(a)=\prod_{i<j}(a_{i}-a_{j})$. One nice property of the integral is 
\begin{equation}\label{i-z}
     \frac{\det||e^{N a_{k}b_{j}}||}{\Delta(a)\Delta(b)}=e^{\frac{N}{2}\Big[\sum_{k=1}^{N}a_{k}^{2}+\sum_{j=1}^{N}b_{j}^{2}\Big]}\frac{\det||e^{-\frac{N}{2}(a_{k}-b_{j})^{2}}||}{\Delta(a)\Delta(b)}.
\end{equation}
Then one can define the quantity
\begin{equation}
    J_{N}(t,A,B)=\frac{1}{t^{\frac{N}{2}}}\frac{\det||e^{-\frac{N}{2t}(a_{k}-b_{j})^{2}}||}{\Delta(a)\Delta(b)}
\end{equation}
which satisfy the partial differential equation
\begin{equation}\label{eq:j}
    2N\frac{\partial J_{N}}{\partial t}=\frac{1}{\Delta(a)}\sum_{k=1}^{N}\frac{\partial^{2}}{\partial a_{k}^{2}}[\Delta(a)J_{N}].
\end{equation}
(\ref{eq:j}) is similar to (\ref{eq:den}) with
\begin{equation}
    U_{k}=\frac{1}{N}\sum\limits_{\substack{j=1\\j\neq k}}^{N}\frac{1}{a_{k}-a_{j}}
\end{equation}
and like (\ref{eq:free}) we can also write 
\begin{equation}
   \lim_{N\to\infty} \frac{1}{N^{2}}\log[J_{N}]=S_{J}[\alpha,\beta,t]-\frac{1}{2}\int\dashint\alpha(a)\alpha(a')\log[a-a']-\frac{1}{2}\int\dashint\beta(b)\beta(b')\log[b-b']
\end{equation}
with $\alpha(a)$ and $\beta(b)$ being the densities of $a$ and $b$. $S_{J}[\alpha,\beta,t]$ satisfies the differential equation
\begin{equation}\label{hjs}
    \frac{\partial S_{J}}{\partial t}=\frac{1}{2}\int\alpha(a)\Big[\Big(\frac{\partial}{\partial a}\frac{\delta S_{J}}{\delta\alpha(a)}\Big)^{2}-\frac{\pi^{2}}{3}\alpha(a)^{2}\Big]da
\end{equation}
one can think of this as the Hamilton-Jacobi equation for the dynamical system with the Hamiltonian
\begin{equation}
    H[\rho,\Pi]=\frac{1}{2}\int\rho(a)\Big[\Big(\frac{\partial}{\partial a}\frac{\delta S_{J}}{\delta\rho(a)}\Big)^{2}-\frac{\pi^{2}}{3}\rho(a)^{2}\Big]da
\end{equation}
We are interested to the solution which connect $\rho(t=0,a)=\alpha(a)$ and $\rho(t=1,b)=\beta(b)$ within time $t=1$.The variational derivative at the end of the trajectory is
\begin{equation}
    \frac{\delta S_{J}}{\delta\alpha(a)}=\Pi(t=0,a),\qquad\qquad \frac{\delta S_{J}}{\delta\beta(b)}=-\Pi(t=1,b).
\end{equation}
The equation of motion which follow from above Hamiltonian can be transform in a single equation (Hopf equation) of the function $f_{J}(t,a)=\frac{\partial \Pi(a)}{\partial a}+i\pi\rho(a)$
\begin{equation}\label{burg}
    \frac{\partial f_{J}}{\partial t}+f_{J}\frac{\partial f_{J}}{\partial a}=0.
\end{equation}
The general solution of (\ref{burg}) can be written in parametric form
\begin{equation*}
\begin{split}
 x=R(\xi)+F(\xi)t, \quad 
f_{J}(t,x)=F(\xi)    .
\end{split}
\end{equation*}
If we introduce two analytic function
\begin{equation*}
    \begin{split}
        G_{+}(x)=x+f_{J}(t=0,x), \quad
        G_{-}(x)=x-f_{J}(t=1,x)
    \end{split}
\end{equation*}
then one can show that 
\begin{equation}
    G_{+}(G_{-}(x))=G_{-}(G_{+}(x))=x.
\end{equation}
After replacing $a_{k}=y_{k}'=\tau y_{k}$ and $b_{j}=\theta_{j}$ followed by $\tau\to i$ we can transform Itzykson-Zuber integral to character of $U_{N}$ group 
\begin{equation}
   \det||e^{N a_{k}b_{j}}||\to  \det||e^{i N y_{k}\theta_{j}}|| =\chi_{R}(U) \Delta(e^{i\theta})
\end{equation}
also the densities of $y'$ and $y$ are related by
\begin{equation}\label{rht}
    \rho_{\tau}(y)=-\frac{\partial x}{\partial y'}=\frac{1}{\tau}\rho(\frac{y}{\tau})|_{\tau=i}.
\end{equation}
With this we can define two type of velocity
\begin{equation}
\begin{split}
U(y) &=\frac{\partial}{\partial y}\frac{\delta S_{J}[\rho_{\tau},\sigma]}{\delta \rho(y)},\quad 
U_{\tau}(y)=\frac{\partial}{\partial y}\frac{\delta S_{J}[\rho_{\tau},\sigma]}{\delta \rho_{\tau}(y)},\\
V(\theta) & =-\frac{\partial}{\partial \theta}\frac{\delta S_{J}[\rho_{\tau},\sigma]}{\delta \sigma(\theta)}, \quad 
v_{2}(\theta)=-\frac{\partial}{\partial y}\frac{\delta S[\sigma_{1},\sigma_{2}]}{\delta \sigma_{2}(\theta)}
\end{split}
\end{equation}
where the quantity $S[\sigma_{1},\sigma_{2}]$ is from (\ref{eq:free}). The above four velocities are related to each other by
\begin{equation}
\begin{split}
     U(y)=\tau U_{\tau}(\tau y)|_{\tau=i},\quad
      -V(\theta)+\theta=-v_{2}(\theta)
\end{split}
\end{equation}
from the large $N$ limit of $\chi_{R}(U)=e^{N^{2}\Xi[\rho,\sigma]}$ and using on shell condition one can show
\begin{equation}
    \frac{\partial}{\partial y}\frac{\delta \Xi[\rho,\sigma_{1}]}{\delta \rho(y)}=\tau U_{\tau}(\tau y)|_{\tau=i}-y=\frac{A}{2}y .
\end{equation}
We can also define two analytic function with these variable
\begin{equation}\label{invg}
    \begin{split}
        G_{+}(x)=x+U_{\tau}(x)+i\pi\rho_{\tau}(x)=-\frac{A}{2}x+-\pi\rho_{\tau}(x)\\
        G_{-}(x)=x-V(x)-i\pi\sigma_{1}(x)=-v_{2}(x)-i\pi\sigma_{1}(x) .
    \end{split}
\end{equation}
Hence from (\ref{eq:in}) with $\sigma_{1}=\sigma_{2}$, in this case $t^{\ast}=\frac{A}{2}$ and at $t=A$ we have
\begin{equation}
    i\pi\sigma^{\ast}[\theta-\frac{A}{2}(v_{2}(\theta)+i\pi\sigma_{1}(\theta))]=v_{2}+i\pi\sigma_{1}(\theta)
\end{equation}
By using (\ref{rht}) and (\ref{invg}) we can show that 
\begin{equation}
    \pi\rho(-\pi\sigma^{\ast}(\theta))=\theta
\end{equation}

\subsection{\texorpdfstring{$Sp(N)$}{SPN}}\label{app:domrepspn}

To find the Young-Tableau density for $Sp(N)$ group we can extend the above procedure of $U(N)$ by a simple observation
\begin{equation}
    \det||\sinh{Na_{k}}b_{j}||=\frac{1}{2^{N}}e^{\frac{N}{2}[\sum_{k=1}^{N}a_{k}^{2}+\sum_{j=1}^{N}b_{j}^{2}]}\det||e^{-\frac{N}{2}(a_{k}-b_{j})^{2}}-e^{-\frac{N}{2}(a_{k}+b_{j})^{2}}||
\end{equation}
helps us to define
\begin{equation}
    J_{N}(t,a,b)=\frac{1}{t^{\frac{N}{2}}}\frac{\det||e^{-\frac{N}{2t}(a_{k}-b_{j})^{2}}-e^{-\frac{N}{2t}(a_{k}+b_{j})^{2}}||}{\Delta(a)\Delta(b)}
\end{equation}
which satisfy the same differential equation as (\ref{eq:j}), with $\Delta(a)=\prod_{i=1}^{N}a_{i}\prod_{j<j}^{N}(a_{i}+a_{j})(a_{i}-a_{j})$. Following the same procedure as depicted in (\ref{sec:spncorrel}) i.e. identifying $a_{-k}=-a_{k}$ and same for $b_{k}$ also,we can write everything in terms of this new $a_{k}$ as
\begin{equation}
    \log[\Delta(a)]=\frac{1}{4}\sum_{i=-N}^{N}\sum\limits_{\substack{j=-N\\j\neq k}}^{N}\log[a_{i}-a_{j}]+\frac{1}{4}\sum_{i=-N}
^{N}\log[2a_{i}]
\end{equation}
with
\begin{equation}
    U_{k}=\frac{1}{N}\log[\Delta(a)]=\frac{1}{N}\sum\limits_{\substack{j=-N\\j\neq k}}^{N}\frac{1}{a_{k}-a_{j}}+\frac{1}{2a_{k}}
\end{equation}
and
\begin{equation}
 \frac{1}{N}\frac{\partial U_{k}}{\partial a_{k}}\approx-4\frac{\pi^{2}}{3}\alpha(a_{k})^{2}   .
\end{equation}
Assuming that $J_{N}$ at the large $N$ is given by $J_{N}=e^{N^{2}F_{N}^{J}}$ and
\begin{equation}
    F_{N}^{J}=S_{N}^{J}-\frac{1}{N^{2}}(\log[\Delta(a)]+\log[\Delta(b)])
\end{equation}
then $S^{J}=\lim_{N\to\infty}S_{N}^{J}$ will satisfy the differential equation
\begin{equation}
    \frac{\partial S^{J}}{\partial t}=\frac{1}{2}\int\alpha(a)\Big[\Big(\frac{1}{2}\frac{\partial}{\partial a}\frac{\delta S^{J}}{\delta\alpha(a)}\Big)^{2}-4\frac{\pi^{2}}{3}\alpha(a)^{2}\Big]da .
\end{equation}
After redefinition of of $S^{J}= \ra 4\tilde{S}^{J}$ the above equation reduces to the Hamilton-Jacobi like equation (\ref{hjs}). Thus the function $f_{J}(t,a)=v(t,a)+i\pi\rho(t,a)$ follows the Burger's equation with the boundary condition
\begin{equation}
    \begin{split}
        Im[f_{J}(t=0,a)]=\pi\alpha(a), \qquad Re[f_{J}(t=0,a)]=\frac{\partial}{\partial a}\frac{\delta {S}^{J}}{\delta\alpha(a)}\\
         Im[f_{J}(t=1,b)]=\pi\beta(b), \qquad Re[f_{J}(t=1,b)]=-\frac{\partial}{\partial b}\frac{\delta {S}^{J}}{\delta\beta(b)}
    \end{split}
\end{equation}
Similar to the $U(N)$ case, after replacing $a_{k}=y_{k}'=\tau y_{k}$ and $b_{j}=\theta_{j}$ followed by $\tau\to i\tau$ we get
\begin{equation}
    \det||\sinh{Ny_{k}'\theta_{j}}||\to\frac{\det||\sin{Ny_{k}\theta_{j}}||}{D[\theta]}(i)^{N}D[\theta].
\end{equation}
We can follow the same procedure as above and prove that when ${\sigma}_{1}(\theta)={\sigma}_{2}(\theta)$ the dominant young tableau density satisfy
\begin{equation}
    \pi{\rho}[-\pi{\sigma}^{\ast}(\theta)]=\theta.
\end{equation}

\bibliographystyle{hieeetr}
\bibliography{Bib}{}

\end{document}